\documentclass[11pt]{article}
\usepackage[margin=1in]{geometry}

\usepackage{times}
\usepackage{graphicx}
\usepackage{amsmath, amsfonts}
\usepackage{url}
\usepackage{hyperref}

\usepackage{booktabs}   
\usepackage{multirow}   

\hypersetup{
    colorlinks=true,
    linkcolor=blue,
    citecolor=blue,
    urlcolor=blue
}

\usepackage{enumitem}
\usepackage{algorithm}
\usepackage{algpseudocode}
\usepackage{amsmath}
\usepackage{xcolor}
\usepackage{subfigure}
\usepackage{multirow}
\usepackage{makecell}   
\usepackage{marginnote}
\makeatletter
\renewcommand{\@biblabel}[1]{\textcolor{blue}{[#1]}}

\definecolor{cgreen}{HTML}{000000} 
\definecolor{cblue}{HTML}{000000}
\definecolor{corange}{HTML}{000000}
\definecolor{cdeepblue}{HTML}{000000}

\newcommand{\rone}[1]{\textcolor{cgreen}{#1}}
\newcommand{\rtwo}[1]{\textcolor{cdeepblue}{#1}}
\newcommand{\rthree}[1]{\textcolor{corange}{#1}}

\newcommand{\rfour}[1]{\textcolor{cblue}{#1}}

\makeatother
\AtBeginDocument{%
  }

\begin{document}

\title{Performant Synchronization in Geo-Distributed Databases}

\author{
Duling Xu\textsuperscript{1}, 
Tong Li\thanks{Tong Li is the corresponding author. Email: \texttt{tong.li@ruc.edu.cn}.}  \textsuperscript{1}, 
Zegang Sun\textsuperscript{1},
Zheng Chen\textsuperscript{2},
\\
Weixing Zhou\textsuperscript{3},
Yanfeng Zhang\textsuperscript{3},
Wei Lu\textsuperscript{1},
Xiaoyong Du\textsuperscript{1}
}

\date{
\textsuperscript{1}Renmin University of China, Beijing, China \\
\textsuperscript{2}Tsinghua University, Beijing, China \\
\textsuperscript{3}Northeastern University, Shenyang, China \\
}

\newcommand{\name}{\texttt{GeoCoCo}}
\newcommand{\api}{\textcolor[RGB]{19,32,67}}
\newcommand{\cred}{\textcolor{red}}
\newcommand{\cblue}{\textcolor{blue}}
\newcommand{\dl}{\textcolor{black}}

\maketitle

\begin{abstract}
The deployment of databases across geographically distributed regions has become increasingly critical for ensuring data reliability and scalability.
Recent studies indicate that distributed databases exhibit significantly higher latency than single-node databases, primarily due to consensus protocols maintaining data consistency across multiple nodes.
We argue that synchronization cost constitutes the primary bottleneck for distributed databases, which is particularly pronounced in wide-area network~(WAN).
Fortunately, we identify opportunities to optimize synchronization costs in real production environments: 1) Network clustering phenomena, 2) Triangle inequality violations in transmission, and 3) Redundant data transfers.
Based on these observations, we propose \name{}, a synchronization acceleration framework for cross-region distributed databases.
First, \name{} presents a group rescheduling strategy that adapts to real-time network conditions to maximize WAN transmission efficiency.
Second, \name{} introduces a task-preserving data filtering method that reduces data volume transmitted over the WAN.
Finally, \name{} develops a consistency-guaranteed transmission framework integrating grouping and pruning.
Extensive evaluations in both trace-driven simulations and real-world deployments demonstrate that \name{} reduces synchronization cost—primarily by lowering WAN bandwidth usage—by up to 40.3\%, and this improves system throughput by up to 14.1\% in GeoGauss.
\end{abstract}

\section{INTRODUCTION}

With the growth of global-scale services, distributed databases have become the standard for managing data across geographically dispersed regions, offering high availability and strong consistency~\cite{corbett2013spanner, thomson2012calvin, geogauss, dang2019towards, hellings2021byshard}.
Yet, maintaining consistency requires frequent synchronization, which introduces significant overhead—especially in geo-distributed settings where transmission can dominate transaction latency (over 60\% in systems like Spanner and GeoGauss)~\cite{corbett2013spanner, zhou2023geogauss}.
These challenges highlight the need for a synchronization-oriented communication framework that reduces coordination complexity and minimizes wide-area communication cost.


Existing efforts to improve synchronization efficiency in geo-distributed databases mainly fall into three categories.
First, transaction-level optimizations mitigate synchronization latency through scheduling and locality-aware execution~\cite{thomson2012calvin, ren2019slog, li2022alnico, choi2023hydra}.
For instance, SLOG~\cite{ren2019slog} pre-orders cross-region transactions near their data to reduce coordination overhead.
Second, network-level approaches accelerate data transfer via advanced transport protocols~\cite{rfc9000, zhang2023quic_fastnet, cardwell2016bbr}.
BBR~\cite{cardwell2016bbr}, for example, estimates bottleneck bandwidth and RTT to maintain high throughput and low latency.
Third, data-reduction techniques compress replicated payloads to lower WAN traffic~\cite{xu2015reducing, shilane2012wan, zlib}.
zlib~\cite{zlib} exemplifies this by providing efficient, lossless compression for synchronization data.

Despite notable performance gains, prior approaches treat synchronization as a coarse-grained monolithic service, missing opportunities for fine-grained hardware utilization. In a fully replicated distributed database, one synchronization round requires $n(n-1)$ point-to-point communications over $n(n-1)$ distinct paths. While intra-datacenter settings assume near-uniform latency (e.g., $\sim$5\,ms) and symmetric bandwidth, this assumption breaks down in geo-distributed environments. For instance, latency within California is under 4\,ms, but rises to 81.12\,ms between Northern California and Central Canada, and 288.45\,ms to Cape Town—spanning two orders of magnitude, with inter-region paths varying by 2--5$\times$. Executing uniform communication over such heterogeneous WAN paths introduces synchronization stalls, motivating a fine-grained re-examination of synchronization bottlenecks.

We conduct empirical measurements and trace-driven analyses to understand synchronization behavior in production-like environments. Our study reveals three key phenomena in geo-distributed deployments: (1) strong geographical aggregation, where certain data centers naturally act as traffic hubs; (2) high redundancy in update propagation, especially among neighboring replicas; and (3) speculative triangular paths that can outperform direct links. These findings indicate that substantial communication overhead can be removed by redesigning synchronization to exploit these real-world characteristics.
Motivated by this, we construct latency-aware synchronization groups aligned with structural properties of geo-distributed systems: (1) identifying natural aggregation hubs for routing and consolidation, (2) eliminating redundant transmissions by leveraging update overlap, and (3) exploiting low-latency triangular paths for faster synchronization. Achieving this requires addressing three challenges. First, group formation and aggregation scheduling must jointly optimize global latency under dynamic and asymmetric WAN conditions. Second, the space of possible grouping strategies grows rapidly with node count, raising scalability concerns. Finally, practical deployment demands a lightweight, modular solution that integrates transparently by modifying only communication invocation logic.

To address these challenges, we propose \name{}, a general-purpose synchronization framework for geo-distributed databases. 
\name{} decouples synchronization from point-to-point communication and instead performs group-based synchronization in two stages: first aggregating updates within groups, then coordinating inter-group transmission via designated aggregators. 
To optimize end-to-end performance, \name{} incorporates a latency-aware group planner and a dynamic aggregation scheduler that jointly minimize makespan. 
In addition, it exposes a high-level interface that enables seamless integration by only modifying communication invocation logic, without altering transaction or storage modules.
We empirically compare \name{} against state-of-the-art baselines in real WAN settings using TPC-C and YCSB. \name{} improves end-to-end throughput by up to 14.1\% on GeoGauss and 11.5\% on CockroachDB, while reducing synchronization (WAN) cost by up to 40.3\%. These results validate \name{}’s effectiveness across systems and workloads. 
In summary, we address the performance bottlenecks of synchronization in geo-distributed databases and make the following 3 \textbf{contributions}:
\begin{itemize}[leftmargin=7pt]
    \item \textbf{\emph{\underline{New Opportunity}:} Fine-Grained Analysis of Synchronization Inefficiencies.} 
    We identify key inefficiencies in geo-distributed synchronization and uncover three actionable opportunities: geographical aggregation hubs, redundant data transmissions, and speculative low-latency triangular paths. 
    These insights lay the foundation for communication optimization beyond traditional designs. 

    \item \textbf{\emph{\underline{New Transmission Schedule}:} Hierarchical Latency-Aware Synchronization.} 
    We propose a hierarchical synchronization design that forms latency-aware groups and routes updates through aggregation hubs, leveraging both structural and speculative transmission advantages. 
    This reduces cross-region overhead and accelerates global coordination.

    \item \textbf{\emph{\underline{New Data Reduction Method}:} Redundancy-Aware Update Dissemination.} 
    We reduce communication volume by identifying overlapping update contents among replicas and proactively removing redundancy through the two-level synchronization hierarchy.
    This redundancy-aware design reduces transmission cost without compromising consistency.
\end{itemize}

\section{BACKGROUND}



\begin{figure}
\centering
\includegraphics[width=0.7\linewidth]{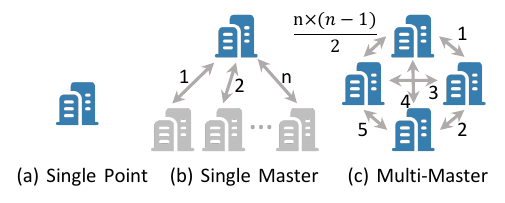}
\vspace{-0.32cm}
\caption{Cross-Domain Communication Complexity in three Database Architectures: (a) Single Point vs. (b) Single Master vs. (c) Multi-Master.}
\label{fig: master}
\vspace{-0.1cm}
\end{figure}

\subsection{Geo-distributed Database System}
\label{sub: geo-dis-db-sys}

To support multinational operations, enterprises deploy data centers across regions, driving the growth of geo-replicated distributed databases. These systems must ensure high availability, strong consistency, and low-latency cross-region access, and are widely adopted in finance, e-commerce, social platforms, and online gaming. The key challenge lies in efficient synchronization and transaction processing over WANs, where high latency, limited bandwidth, and asynchrony are inherent.
With the expansion of cloud and edge computing, demand for scalable and flexible data services continues to rise, accelerating the evolution of geo-replicated databases~\cite{corbett2013spanner,taft2020cockroachdb,verbitski2017amazonaurora,zhou2023geogauss,cao2022polardb}. Representative systems such as Google Spanner~\cite{corbett2013spanner}, CockroachDB~\cite{taft2020cockroachdb}, Amazon Aurora~\cite{verbitski2017amazonaurora}, and GeoGauss~\cite{zhou2023geogauss} have advanced consensus, replication, and network-aware optimizations, establishing geo-replicated databases as critical infrastructure for enterprise competitiveness.
Mainstream systems adopt either \textbf{Single-Master} or \textbf{Multi-Master} architectures (Figure~\ref{fig: master}). In Single-Master systems (e.g., Spanner~\cite{corbett2013spanner}, TiDB~\cite{huang2020tidb}, CockroachDB~\cite{taft2020cockroachdb}), leaders handle writes and replicate to followers, incurring high cross-region latency due to centralized routing and 2PC overhead~\cite{gray2005notes}. In contrast, Multi-Master systems eliminate the single-leader bottleneck but introduce $O(n^2)$ synchronization cost, as each update requires $n(n-1)$ cross-region messages~\cite{zhou2023geogauss}. Consequently, communication scalability remains the primary barrier to global-scale deployment.

\begin{figure}
    \centering
    \includegraphics[width=0.7\linewidth]{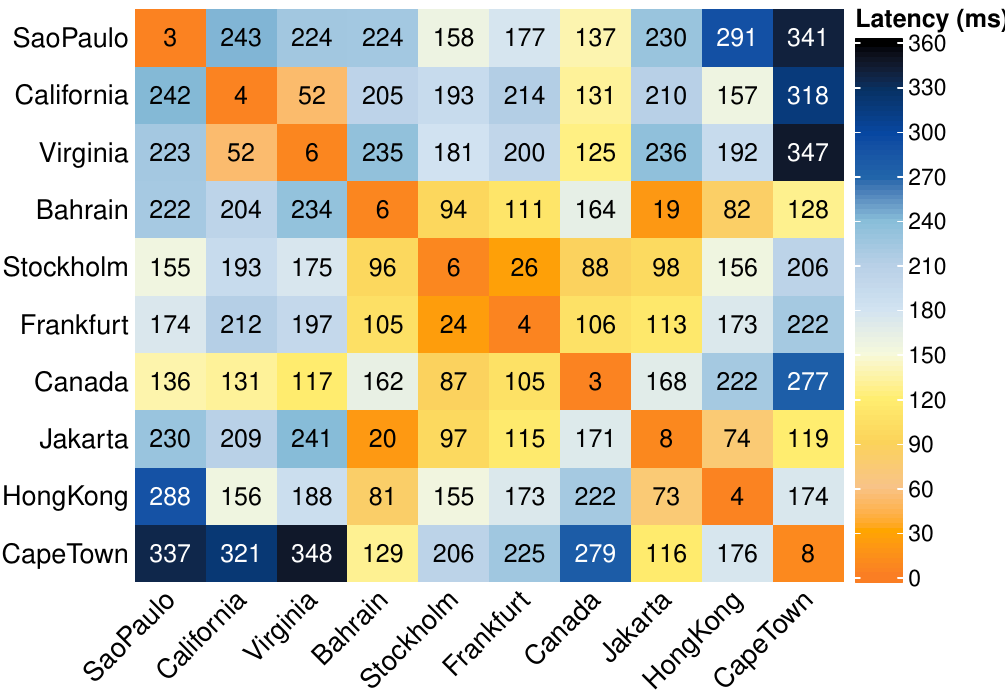}
    \vspace{0.2cm}
    \caption{Measured network latency between Amazon EC2 sites in 10 different regions.}
    \label{fig: latency-differ}
\vspace{-0.01cm}
\end{figure}
    
\subsection{WAN Bandwidth and Latency}
\label{sub: wan-bandwidth-latency}

The network characteristics of Wide Area Network (WAN) differ fundamentally from those of intra-datacenter Local Area Networks (LANs), posing significant challenges to the design of geo-distributed systems. On modern cloud platforms, WAN environments exhibit three critical limitations: limited bandwidth, high and variable latency, and unpredictable latency fluctuations~\cite{LI2025371}.
\textbf{First}, WAN bandwidth is significantly lower than LAN bandwidth—on average 15× and up to 60–80× lower in extreme cases~\cite{hsieh2017gaia, yao2023ragraph}. Purchasing WAN bandwidth comparable to LAN speeds is also substantially more expensive, often 10× higher per unit~\cite{aws-ec2-pricing, kumar2021aggnet}. These cost-performance trade-offs make large-scale, bandwidth-heavy tasks (e.g., all-reduce, global synchronization) highly constrained under WAN settings~\cite{pu2015low}.
\textbf{Second,} WAN suffer from high absolute latency and inter-region variability(see Figure~\ref{fig: latency-differ}). Measurements from 10 global regions on Amazon’s cloud~\cite{aws_network_manager_infra_perf} show latencies ranging from 26 ms (Stockholm–Frankfurt) to over 337 ms (São Paulo–Cape Town), introducing nearly 10× variability. Latency jitter further exacerbates coordination unpredictability. In Multi-Master databases, even a single region’s delay can postpone global snapshots and validation, stalling progress across all sites.
In short, WAN introduces severe performance bottlenecks that undermine the efficiency of geo-distributed coordination. Addressing these bandwidth and latency constraints forms the basis for our system’s design and analysis in subsequent sections.

\begin{figure}[t]
\centering
    \subfigure[  Varying Bandwidth.]{\label{fig: width}
    \begin{minipage}[]{0.38\linewidth}
        \includegraphics[width=\linewidth]{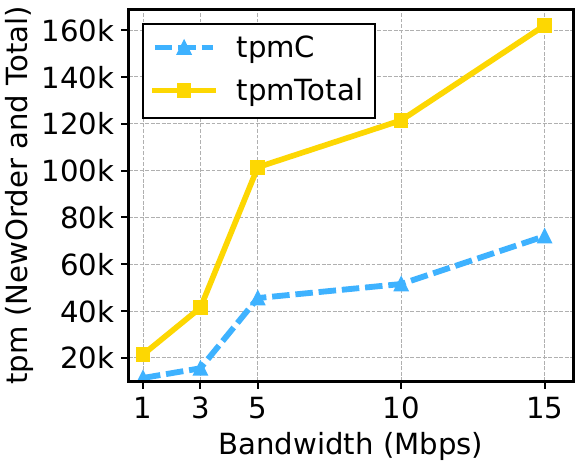}
    \end{minipage}
    }
    \subfigure[  Varying Latency.]{
        \label{fig: latency}
        \begin{minipage}[]{0.38\linewidth}
            \includegraphics[width=\linewidth]{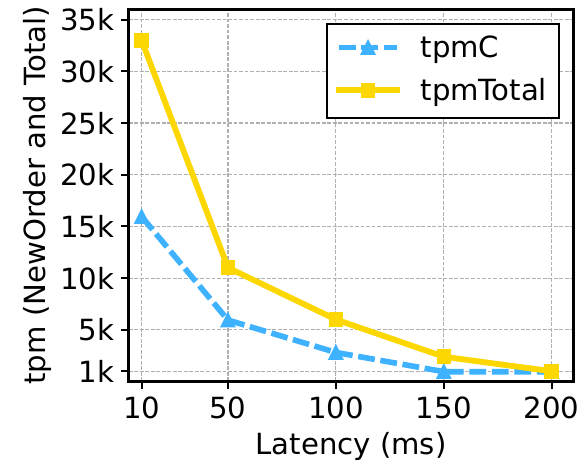}
        \end{minipage}
    }
\vspace{-0.4cm}   
\caption{Performance Impact of WAN Bandwidth and Latency on Multi-Master Systems Transactions.} 
\label{fig: bac-2}
\vspace{-.2in} 
\end{figure}

\vspace{-.1in}                        
\subsection{Distributed Database System on WAN}
\label{sub: dis-sb-sys-on-wan}
Geographically distributed Multi-Master databases face inherent WAN limitations. While they enable low-latency local access, maintaining global consistency incurs high WAN overhead, as every update must reach remote replicas. We evaluate this effect on GeoGauss using a five-node geo-distributed setup (Figure~\ref{fig: bac-2}) running the TPC-C workload~\cite{tpcc}.
Reducing bandwidth from 15 Mbps to 1 Mbps rapidly saturates the network and degrades throughput, as each transaction’s writes are broadcast to all replicas. Likewise, increasing RTT from 10 ms to 200 ms significantly raises latency and lowers throughput, since strong consistency requires acknowledgments from all sites~\cite{corbett2013spanner, Li2012RedBlue}.
These results confirm that WAN latency and bandwidth dominate system performance, shifting the bottleneck from computation to communication and motivating our WAN-aware synchronization design.

\textbf{In this work, we focus on abstracting these collective communication patterns over WAN,} as geo-distributed multi-master databases require a collective communication framework that operates beyond the boundaries of a single data center.

\section{MOTIVATION}





After conducting an in-depth exploration of distributed databases deployed in real-world wide area networks (WANs), we present three key observations.

\begin{figure}
\vspace{0.2cm}
    \centering
    \begin{minipage}{0.4\linewidth}
        \includegraphics[width=\textwidth, height=125pt]{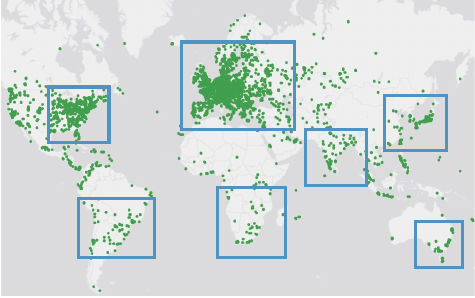}
        \vspace{-0.2cm}
        \caption{Observed Geographical Clustering of Data Centers Worldwide.}
        \label{fig: cluster}
    \end{minipage}
    \hspace{0.04\linewidth}
    \begin{minipage}{0.4\linewidth}
        \includegraphics[width=\textwidth,height=125pt]{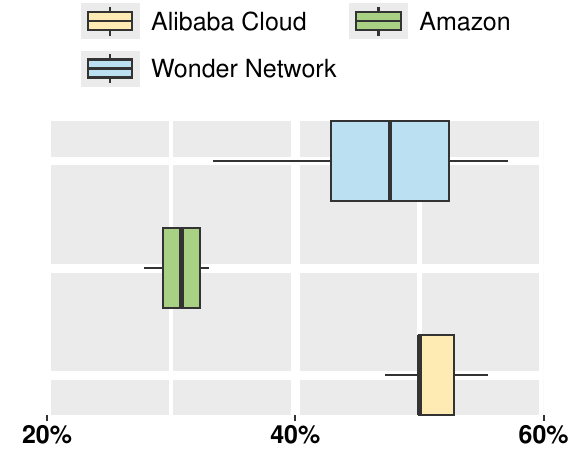}
        \vspace{-0.2cm}
        \caption{Proportion of nodes that violate the triangle inequality~(\%).}
        \label{fig: trangle}
    \end{minipage}
\vspace{-0.5cm}
\end{figure}


\vspace{0.3ex} 
\noindent 
\textbf{Observation \#1: Geographically-close nodes naturally form low-latency clusters.}
\noindent Due to geographic proximity, data centers naturally form small clusters(see Figure~\ref{fig: cluster}). Nodes within the same cluster experience significantly lower communication latency, whereas inter-cluster communication exhibits considerably higher delays. This characteristic provides an optimization opportunity for layered communication strategies: prioritizing intra-cluster communication and treating inter-cluster communication as an independent layer for optimization. For example, by minimizing the use of high-latency inter-cluster paths, the latency bottleneck can be effectively alleviated, and more refined group-based communication management becomes feasible.


\vspace{0.3ex}
\noindent 
\textbf{Observation \#2:~At high conflict ratios, some WAN traffic becomes wasteful and ineffective.}
We define \emph{white data} as updates that are eventually discarded during synchronization without affecting the final system state. 
In geo-distributed multi-master databases, such data often stems from conflicting writes, duplicate updates, or stale transactions~\cite{geogauss, thomson2012calvin, wu2016transaction}. Although white data consumes considerable WAN bandwidth—especially under all-to-all patterns—it holds no semantic value and is ultimately filtered by the application layer~\cite{10.14778/3342263.3342627}. 
Eliminating such data reduces WAN overhead without compromising consistency and forms a core optimization target in our design.
To quantify its impact, we measured white data ratios in a geo-distributed database. 
\rfour{
Depending on workload and conflict intensity, it accounts for 20.12–45.23\% of transmitted updates in our TPC-C and YCSB experiment, introducing unnecessary bandwidth usage and queuing delay~\cite{song2022swan}.
In our experience with Huawei’s production environments, up to 20–45\% of synchronized updates are “white data” with no observable remote effect, demonstrating filtering’s practical value under standard benchmarks. 
}
Early filtering reduces total transmission cost and improves end-to-end latency. When combined with semantic aggregation at intermediate nodes, communication becomes significantly more efficient.

\vspace{0.3ex} 
\noindent 
\textbf{Observation \#3:~The link latencies in wide-area networks often violate the triangle inequality.}
\noindent In data center networks, the triangle inequality typically holds, meaning that the sum of two latency segments is always greater than the direct latency of a single hop. However, in WAN, this rule is frequently violated, we call it Triangle Inequality Violation (TIV). For instance, the direct communication latency between nodes A and C may be 100ms, whereas routing through an intermediate node B (A → B → C) results in a total latency of only 80ms. This forms a triangle in which the latency inequality is violated. 
As shown in Figure~\ref{fig: trangle}, statistical data from 3 real-world WAN datasets indicate that 28~\%-57~\% of node pairs exhibit such violations~\cite{alicloud_nis_inter_region, aws_network_manager_infra_perf, wondernetwork}. 
This counterintuitive phenomenon creates an opportunity for latency optimization: by dynamically selecting indirect paths when they offer lower delay, communication latency—and thus makespan—can be reduced. This is especially beneficial for latency-sensitive operations such as all-reduce and all-to-all.
\vspace{-0.1cm}

\section{\name{} DESIGN}

\subsection{\name~ Overview}
\label{sub: overview}

\begin{figure}
    \centering
    \includegraphics[width=0.7\linewidth]{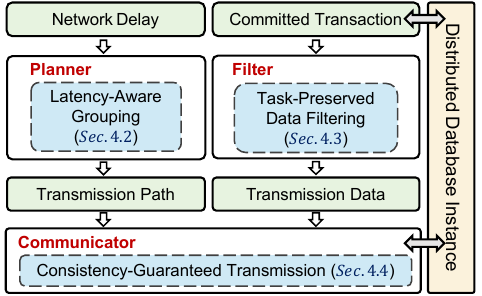}
    \vspace{-0.1cm}
    \caption{Overview of GeoCoCo.}
    \label{fig: overview}
\vspace{-0.4cm}
\end{figure}


\name{} is a distributed transmission framework designed to accelerate synchronization in geo-distributed database systems by optimizing collective communication, thereby enhancing overall database performance.
As illustrated in Figure~\ref{fig: overview}, \name{} is deployed between the communication subsystem and distributed database instances, acting as an intelligent adaptation framework that dynamically optimizes cross-domain data flows.
The system consists of three core modules: the Planner, the Filter, and the Communicator, each responsible for distinct dimensions of optimization.

\noindent 
\textbf{\underline{Planner}}. The Planner module is responsible for optimizing transmission coordination through real-time WAN monitoring and adaptive participant grouping. Observation~\#1 inspires the design of \name{}’s grouped transmission mechanism. The Planner continuously monitors wide-area network conditions—such as link latency—via a built-in real-time observation component. Based on the observed network dynamics, it invokes a linear programming–based grouping strategy orchestrator that adaptively clusters transmission participants to balance latency and resource utilization. By intelligently forming sender groups, this mechanism effectively reduces the number of WAN round-trips required per transmission round and collaborates with the Communicator module to suppress redundant traffic through selective filtering.

\noindent
\textbf{\underline{Filter}}.
This component filters out “white data”—updates irrelevant to the receiver’s current task context—thereby reducing unnecessary communication overhead in geo-distributed environments. By analyzing task dependencies and application semantics, the filter selectively transmits only data that contributes to the current transaction, minimizing bandwidth usage and alleviating the burden on downstream consistency mechanisms. Observation~\#2 motivates the design of \name{}’s filtering module, with the grouping plan providing essential conditions for identifying irrelevant data.

\noindent 
\textbf{\underline{Communicator}.}
The Communicator module is responsible for coordinating the final stage of transmission. It receives the optimized transmission path plan from the \textit{Planner} and the filtered data from the \textit{Filter} module, and accordingly controls the actual data delivery process.
The Communicator integrates a consistency-preserving transmission mechanism that ensures the pruned data still meets the consistency requirements of the distributed database system. By decoupling transmission coordination and data filtering from the final delivery mechanism, this module enables efficient and reliable synchronization across geo-distributed nodes without requiring additional coordination.
The hierarchical design in Communicator is jointly inspired by Observation~\#1 and Observation~\#3.

\noindent
\rtwo{
\underline{\textbf{Novelties.}}
To our knowledge, GeoCoCo is the first system that systematically optimizes all-to-all synchronization in geo-distributed transactional databases.
It introduces three key techniques:
(1) a \textbf{latency-aware clustering framework} for topology-adaptive grouping;
(2) a \textbf{semantic filtering mechanism} that removes “white data” to reduce redundant communication; and
(3) a \textbf{group-based transmission mechanism} enabling efficient aggregated synchronization across groups.
Together, these techniques jointly enhance the structural, semantic, and execution efficiency of synchronization.
\vspace{-0.3cm}
}

\subsection{Latency-aware Grouping Strategy Orchestrator}
\label{subsec: grouping}

In this section, we introduce the process of generating the group plan, including real-time latency monitoring and a latency-aware grouping strategy.


\noindent
\textbf{Real-Time Latency Monitoring.}
We conduct real-time monitoring of inter-node latency in the distributed system.
This component dynamically produces an $N \times N$ latency matrix $L$, where each entry $L[i,m]$ represents the current measured latency between node $i$ and node $m$. 
This latency matrix is updated in real-time to reflect changing WAN conditions, and it serves as the foundational input to our grouping strategy. 
By feeding fresh latency data into the planning algorithm, \name{} ensures that grouping decisions remain adaptive to the network’s state.

\noindent
\textbf{Design Goal.} 
Given the latency matrix $L$ and a desired number of groups $k$, our goal is to partition the $N$ nodes into $k$ communication groups and select one aggregator node for each group, so as to minimize the overall communication delay $T$ across the system.

\vspace{1ex}
\noindent
\textbf{Problem Formulation.} 
We formalize the latency-aware grouping optimization problem with the following variables and objective.

\vspace{1ex}
\noindent \textbf{1. Decision Variables.}
We define binary variables $x_{i,j}$ and $y_{i,j}$ to represent grouping and aggregator selection: $x_{i,j}=1$ if node $i$ belongs to group $j$, and $y_{i,j}=1$ if node $i$ is the aggregator of group $j$. Each node is assigned to exactly one group, and each group has exactly one aggregator. We further introduce continuous auxiliary variables to capture latency costs, where $l_j$ denotes the maximum intra-group latency of group $j$, and $L$ denotes the total inter-group latency.

\vspace{1ex} 
\noindent \textbf{2. Objective Function.} 
Based on the aforementioned variables, we define the total cost of a single synchronization round as follows:
\begin{equation}
        T = max(l_j) + max(L)
\end{equation}
where
\begin{equation}
        L_j = max(L[k, j]), \forall x_{i_j} = 1 
\end{equation}
and
\begin{equation}
    L = max(L[u, v]), \forall y[u, u] = 1 \land y[v, v] = 1 \land u \neq v
\end{equation}


\begin{algorithm}[t]
\caption{Latency-Aware Grouping via LP Optimization}
\label{alg: lp-grouping}


\begin{flushleft}
\textbf{Require:} Latency matrix $\mathcal{L}$ ($N \times N$), number of groups $k$ \\
\textbf{Ensure:} Optimal group assignment and aggregator selection
\end{flushleft}

\begin{algorithmic}[1]

\State \textbf{\textcolor{blue}{Initialize LP Objective}}
\State \quad Minimize total communication delay $T$

\State \textbf{\textcolor{blue}{Define Variables}}
\State \quad $x_{i,j}$: Node $i$ belongs to group $j$
\State \quad $y_{i,j}$: Node $i$ is the aggregator of group $j$
\State \quad $l_j$: max intra-group delay for group $j$
\State \quad $L$: total inter-group delay; $T$: overall delay

\State \textbf{\textcolor{blue}{Add Constraints}}
\For{each node $i$}
    \State $\sum\limits_{j=1}^{k} x_{i,j} = 1$ \Comment{node must join one group}
\EndFor
\For{each group $j$}
    \State $\sum\limits_{i=1}^{N} y_{i,j} = 1$ \Comment{one aggregator per group}
\EndFor
\State $y_{i,j} \le x_{i,j}$ \Comment{Only group members can be aggregators}
\State Define $z_{i,m,j}$: node $i$, $m$ in group $j$
\State Define $w_{i,m,j_1,j_2}$: cross-group pairs
\State $l_j \ge \mathcal{L}[i,m] \cdot z_{i,m,j}$
\State $L \ge \sum \mathcal{L}[i,m] \cdot w_{i,m,j_1,j_2}$
\State Ensure $T \ge \max_j l_j$ and $T \ge L$

\State \textbf{\textcolor{blue}{Solve and Return}}
\State Use LP solver to minimize $T$
\State Extract $x_{i,j}$ and $y_{i,j}$ to get $\textit{group\_plan}$
\State \Return $\textit{group\_plan}$
\end{algorithmic}
\end{algorithm}
\vspace{-0.4cm}




\vspace{1.4ex} 
\noindent \textbf{3. Constraints.} 
To correctly model the grouping problem, we impose a set of linear constraints that ensures a valid grouping and binds the $T$ value to the latency measures.
With the above formulation in place, we leverage a linear programming solver (more precisely, a mixed-integer linear program, given the binary variables) to determine the optimal grouping. 
Algorithm~\ref{alg: lp-grouping} outlines this latency-aware grouping procedure. 
In summary, we initialize the solver to minimize $T$ (Lines 1–2) and define variables~(Lines 3-7).
Then we add all the constraints (Lines 9–20). 
We impose three constraints. 
First, each node must join exactly one group (Lines 9–11), ensuring no node is unassigned or duplicated. 
Second, each group must have exactly one aggregator (Lines 12–14). 
Third, a node may serve as a group's aggregator only if it belongs to that group (Line 14). 
Solving the model yields the optimal $x_{i,j}$ and $y_{i,j}$ values (Lines 22–24), specifying group membership and aggregator selection, and producing a plan that minimizes latency cost $T$.

This latency-aware orchestrator is invoked periodically or when network conditions change, ensuring that groups adapt to maintain low-latency communication. 
\rfour{
In practice, wide-area network dynamics are typically episodic rather than continuous, exhibiting diurnal demand patterns, bursty anomalies, and event-driven routing changes~\cite{hong2013achieving, yao2023ragraph}.
As a result, stable communication phases often last long enough to amortize the cost of grouping decisions.
\name{} does not rely on per-packet or per-millisecond reconfiguration; instead, it performs periodic or event-driven plan updates, avoiding excessive control overhead while still tracking meaningful network changes.
To further prevent oscillation caused by transient RTT noise, \name{} adopts a Re-group damping strategy: re-grouping is triggered only under sustained latency deviation (e.g., >20\% over a sliding window), while unnecessary reconfigurations are suppressed.
}
The grouping strategy is crucial for enabling hierarchical transmission (where aggregators forward data among themselves in a second tier of communication, see subsection~\ref{subsec: trans}) and for providing a structured basis on which task-level data filtering can operate efficiently (subsection~\ref{subsec: filtering}). 
In essence, by clustering nodes intelligently, \name{} reduces costly WAN round-trips and confines most traffic to low-latency local groups, substantially lowering both intra- and inter-data-center communication costs.

\vspace{0.3ex}
\noindent
\textbf{\name{} in Larger Cluster.} 
In practical geo-distributed database deployments, each logical node typically corresponds to a data center, and the number of such nodes is usually limited. Real-world cross-region systems commonly involve up to around 25 data center nodes~\cite{charapko2021pigpaxos}. This aligns well with the practical range of group numbers k considered in our model.
However, to ensure scalability and generality, we also evaluate scenarios with larger node counts to explore future expansion and stress-test the planner’s performance in \S \ref{sub: deep-dive}.
Moreover, considering the computational cost of LP-based planning, we conduct a theoretical analysis of the optimal group number $k^*$ to guide the search space and improve planner efficiency under different node scales.

\vspace{0.5ex}
\noindent
\textbf{The Setting of Group Number.}
As the group size increases (i.e., as the number of nodes \( N \) grows), the computation time of LP-based grouping also increases accordingly. 
To optimize the overall completion time in practical systems, 
our goal is to determine an appropriate number of communication groups \( k \). 
Although the actual completion time may also be affected by factors such as bandwidth limitations, network congestion, and dynamic path selection, communication latency typically remains the dominant performance bottleneck. 
Therefore, selecting an appropriate \( k \) to balance computational overhead and communication efficiency is crucial for performance optimization.
To guide the design space, we adopt a refined cost model of the form of communication cost:
\begin{equation}
    C_{\text{total}} = 2N\left(\frac{N}{k} - 1\right) + 2k(k - 1)
\end{equation}

\begin{equation}
    k^* = \left( \frac{N^2}{2} \right)^{1/3}
    \label{equation: k}
\end{equation}

which more accurately reflects the total communication load under a hierarchical all-to-all scheme. The first term represents the cumulative intra-group cost across $k$ groups, assuming each group of size $n = \frac{N}{k}$ performs full pairwise exchange. The second term models inter-group communication cost among $k$ aggregators.
By analyzing the derivative of the model to identify its minimum, we find that the total communication cost is minimized when the number of groups satisfies Equation(\ref{equation: k}).
The exact value of the optimal group count $k^*$ varies with the cluster size $N$. For smaller clusters ($N \leq 25$), $k^*$ typically falls within the range of $[N/3, N/2]$. As $N$ increases, the ratio $k^*/N$ decreases rapidly. Therefore, we narrow the search range around $k^*$ with a small tolerance. This strategy enables efficient and accurate group selection across different cluster scales, and we validate its effectiveness in \S\ref{sub: group-number}.
\vspace{-0.1cm}

\subsection{Task-Preserved Data Filtering}
\label{subsec: filtering}

\name{}’s filtering mechanism is integrated into its hierarchical communication process, using contextual metadata about each update to decide its relevance. In a multi-master database scenario (such as the GeoGauss~\cite{geogauss}), each replica executes transactions locally using OCC~\cite{geogauss, wang2016mostly,kraska2009consistency} and periodically exchanges batched updates with other replicas. \name{} leverages metadata inherent in this process—e.g., transaction identifiers, read/write sets, timestamps, and update payloads—to perform rule-based filtering that prunes out irrelevant data before it leaves the local region.

\vspace{0.5ex}
\noindent
\textbf{Definition and Impact of White Data.}
In geo-distributed systems, \emph{white data} refers to updates that are transmitted but ultimately discarded during synchronization, contributing nothing to the final state of the receiving replica. 
Filtering white data—defined earlier—requires balancing trade-offs between transmission reduction, correctness, and computational overhead.
\textbf{Redundant content}: semantically identical updates repeatedly sent;  
\textbf{Conflicting or stale updates}: caused by asynchronous execution or validation failures;  \textbf{Null or sparse data}: updates with no meaningful payload.


\noindent While white data consumes WAN bandwidth and introduces queuing overhead, it has no effect on system correctness or transactional integrity. Eliminating such data reduces communication cost and improves latency. This concept, generalized from our observations in Section 3, forms the foundation of \name{}'s filtering design. By identifying and removing white data early, \name{} achieves lossless filtering: the system’s externally visible behavior (e.g., commit results or model convergence) remains unchanged, but with significantly reduced WAN traffic.

\begin{figure}[t]
\vspace{-0.1cm}
    \centering
    \includegraphics[width=0.68\linewidth]{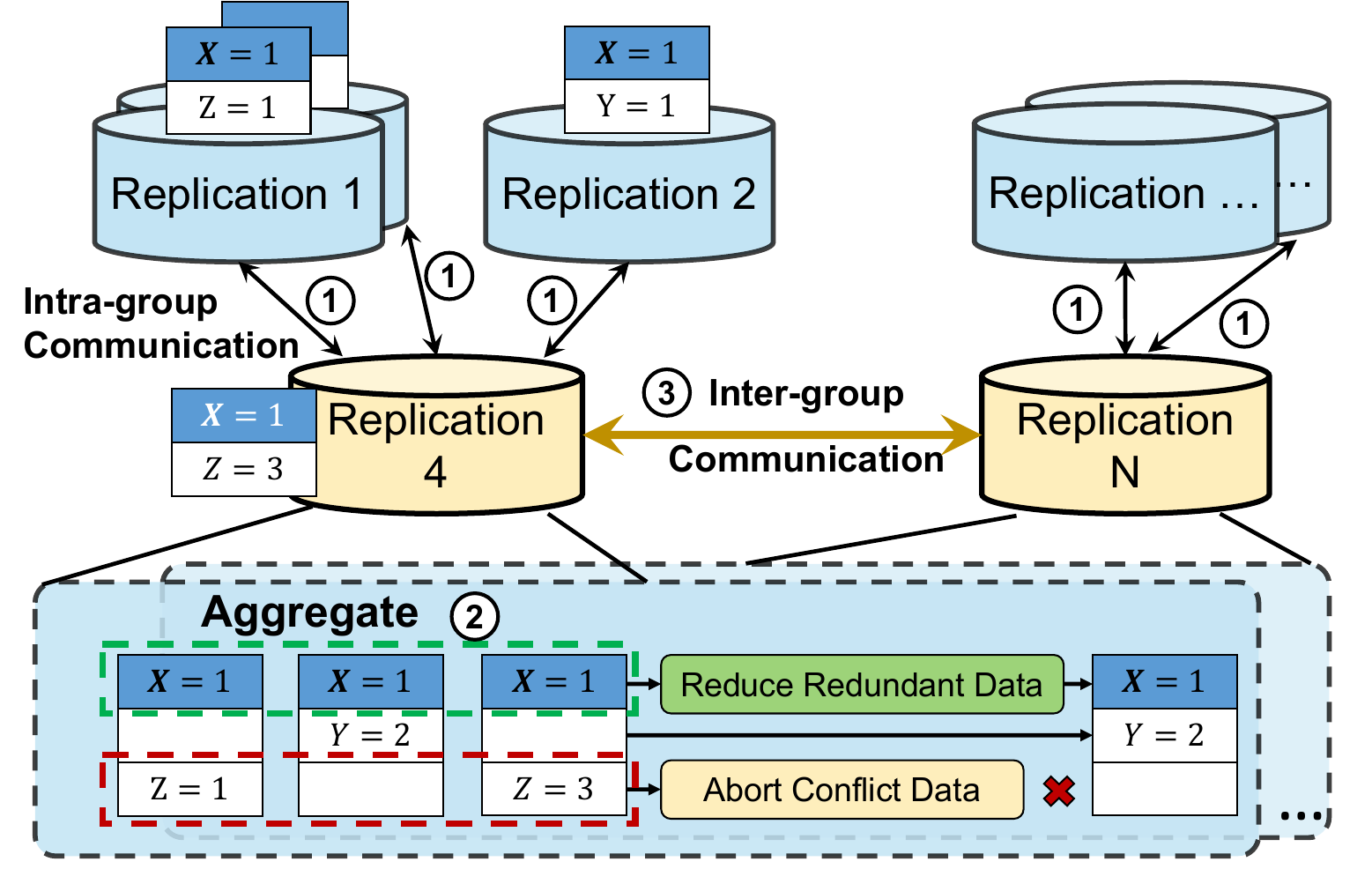}
    \vspace{-0.1cm}
    \caption{The filtering process during synchronization in geo-distributed multi-master databases (e.g., Geo-Gauss synchronization).}
    \label{fig: fig-filter}
\vspace{-0.3cm}
\end{figure}


\vspace{0.4ex}
\noindent
\textbf{Design of the Task-Preserved Filtering Mechanism.}
We illustrate the white data filtering process using a multi-master database as an example. In such systems, write operations are typically handled via OCC, where transactions execute locally and periodically exchange batched updates across replicas (e.g., GeoGauss). The filtering workflow is depicted in Figure~\ref{fig: fig-filter}, assuming group members and leaders have already been determined:
(1) Intra-group transmission. Each member node sends local information (e.g., read/write sets) to its designated leader. If a group consists only of a leader, it proceeds directly to the next step.
(2) Information aggregation. The leader aggregates metadata 
from its members and applies rule-based screening. 
In database settings, this first includes conventional conflict detection based on read/write-set validation, where conflicting transactions are aborted before cross-group propagation. 
In addition, \name{} identifies white data---updates that do not change the visible database state---and filters them out to avoid unnecessary WAN transmission. 
The filtering logic performs constant-time version-vector and hash checks over local metadata at the aggregation node, avoiding global coordination and ensuring $O(1)$ per-update overhead.
Because the checks are entirely local and metadata footprint is bounded, the filtering cost avoids blow-up as the system scales, ensuring it does not become a bottleneck even in large clusters.
(3) Inter-group commit and broadcast. The leader commits the filtered and aggregated results across groups and disseminates the final data to all group members.

In summary, \name{}'s filtering framework integrates two complementary strategies. 
First, it performs hierarchical path optimization by leveraging the Triangle Inequality through multi-hop grouping, enabling speculative yet low-latency synchronization across regions. 
Second, it applies content-aware filtering that proactively detects and eliminates conflicting or irrelevant updates—such as aborted transactions and white data—early in the transmission process. 
This ensures that only valid, task-preserving operations are propagated through the hierarchy, significantly reducing bandwidth consumption and improving end-to-end latency.

    
    

    
    

\subsection{Consistency-Guaranteed Transmission}
\label{subsec: trans}

\begin{figure}[!t]
  \centering
  \includegraphics[width=0.7\linewidth]{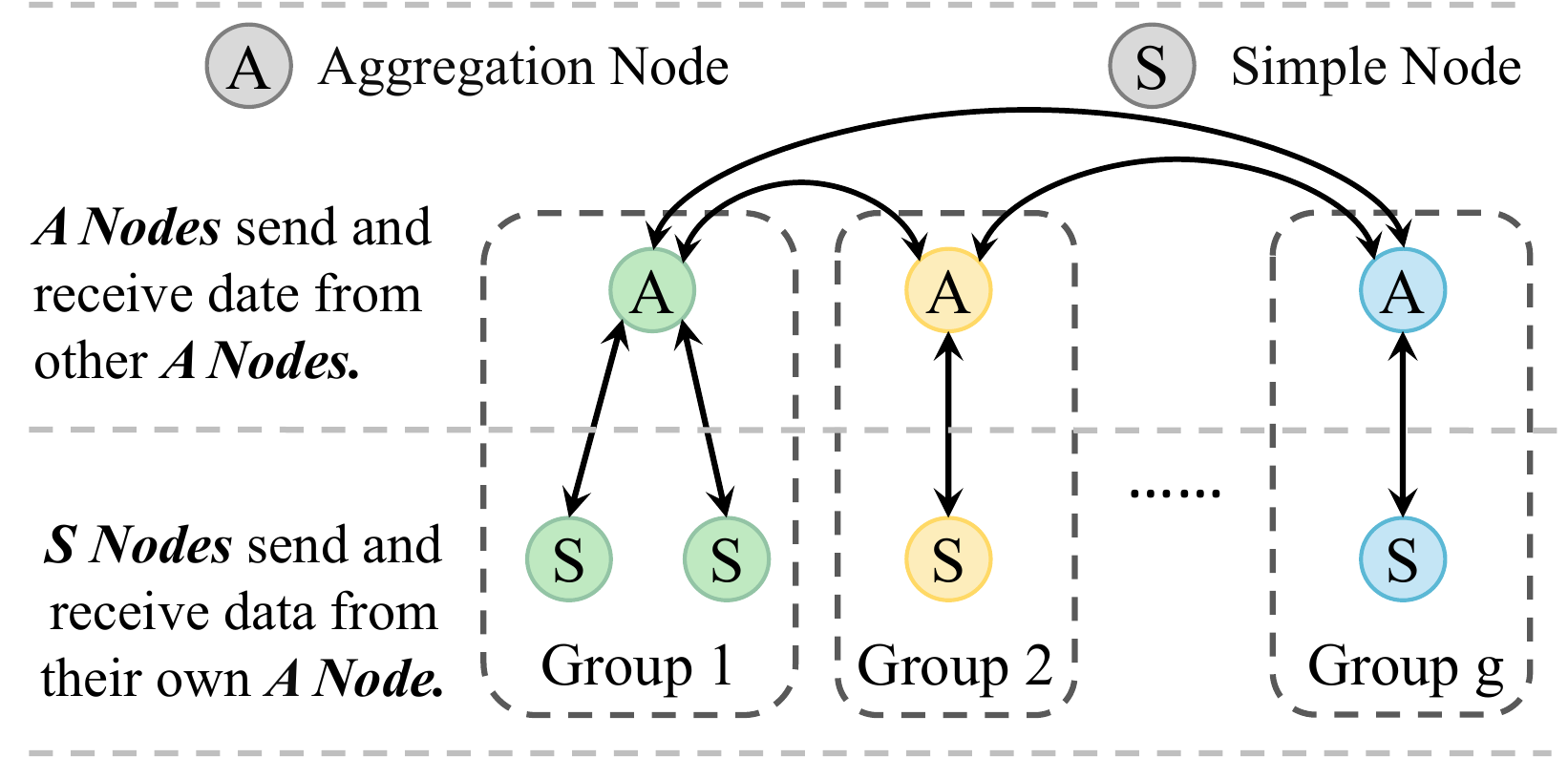}
  \vspace{-0.1cm}
  \caption{Hierarchical transmission.}
  \label{fig: hierarchical}
\end{figure}

\vspace{0.3ex}
\noindent
\textbf{Hierarchical Transmission Flow.}
\name{} introduces a hierarchical, consistency-preserving transmission mechanism that optimizes WAN data flow without altering the underlying consistency model. Nodes are dynamically partitioned into latency-aware groups via an LP-based planner, and one \textit{Aggregation node (A)} is selected per group (Figure~\ref{fig: hierarchical}). Each \textit{Simple node (S)} sends updates only to its local A, which aggregates intra-group data (with filtering; see \S\ref{subsec: filtering}) and forwards consolidated results to other aggregation nodes. After inter-group exchange, each A disseminates received updates back to its group.
This design clusters low-latency peers and minimizes long-haul WAN edges, significantly reducing synchronization cost while preserving correctness. Aggregators adapt across rounds based on network conditions, ensuring efficiency under dynamic WAN behavior. Importantly, S nodes never communicate cross-group in a round—A nodes orchestrate all inter-group transfer—allowing \name{} to reshape communication paths while fully retaining the system’s original consistency semantics.

\vspace{0.3ex}
\noindent
\rthree{
\textbf{Correctness under Message Reordering, 
Delayed Updates, and Network Partition.}
\name{} inherits the correctness guarantees of GeoGauss’s epoch-aware delta-CRDT replication model, which provides strong convergence even under unreliable communication~\cite{geogauss, shapiro2011crdt}. 
\textbf{First}, the CRDT merge function $\oplus$ satisfies three key algebraic properties: 
\emph{Commutativity}, \emph{Associativity}, and \emph{Idempotence} (ACI). 
Let $S$ be the current system state and 
\(
\mathcal{U} = \{u_1, u_2, \dots, u_m\}
\)
the set of updates produced in an epoch.
If all updates are applied exactly once in any order, the merged state is
\[
S' = S \oplus u_1 \oplus u_2 \oplus \dots \oplus u_m.
\]
More generally, consider any permutation $\pi$ of the update set and any multiplicity vector
\(\mathbf{k} = (k_1, k_2, \dots, k_m)\),
where $k_i$ indicates how many times $u_i$ is delivered (including duplicates). 
The resulting state after arbitrary reordering and duplication is
\[
S' = S \oplus
\underbrace{u_{\pi(1)} \oplus \dots \oplus u_{\pi(1)}}_{k_{\pi(1)}\ \text{times}}
\oplus \dots \oplus
\underbrace{u_{\pi(m)} \oplus \dots \oplus u_{\pi(m)}}_{k_{\pi(m)}\ \text{times}}.
\]
Because $\oplus$ is \textbf{commutative and associative}, the order of updates does not affect the result; 
and because it is \textbf{idempotent} ($x \oplus x = x$), repeated applications of the same update have no effect. 
Thus, for any $\pi$ and $\mathbf{k}$:
\[
S \oplus \bigoplus_{i=1}^{m} \bigoplus_{j=1}^{k_i} u_i
= S \oplus u_1 \oplus u_2 \oplus \dots \oplus u_m.
\]
Intuitively, $\oplus$ behaves like a set union: duplicates collapse to one, and the final result depends only on the set of updates, not their order or number of deliveries. 
This algebraic property is the foundation for \name{}’s correctness under message reordering, duplication, and delayed delivery.
\textbf{Second}, \name{} strictly enforces \emph{epoch boundaries}, preventing cross-round reordering. 
Delayed updates that miss epoch $e$ are incorporated into epoch $e+1$, delaying visibility but not affecting correctness or convergence. 
\textbf{Third}, under message delay, the additional visibility latency is bounded. 
If $\tau$ denotes the retransmission timeout and $\Delta_{\mathrm{WAN}}$ denotes the maximum WAN propagation delay, the \textbf{worst-case additional visibility delay} is:
\[
T_{\text{delay}}^{\max} \leq \tau + \Delta_{\mathrm{WAN}},
\]
after which the delayed update is guaranteed to be merged in the subsequent epoch.
\textbf{Finally}, in the presence of \emph{network partitions}, \name{} preserves \emph{safety} but may temporarily reduce liveness. 
Because GeoGauss generates an epoch snapshot only after receiving and merging all updates from the epoch, cross-partition updates are buffered and not committed until the partition heals. 
This prevents divergent states: after reconciliation, all replicas deterministically converge to the same state as if no partition had occurred. 
Thus, network partitions may increase visibility latency but cannot violate correctness.
\textbf{In summary}, \name{} preserves end-to-end application-level consistency under arbitrary message reordering, delivery delays, and network partitions. 
Reordering and delay are fully absorbed by CRDT algebraic properties and epoch isolation, while partitions affect only progress, not correctness.
}

\vspace{0.3ex}
\noindent
\textbf{Transmission Round Guarantee.}
We guarantee that the number of communication rounds required for synchronization under hierarchical transmission is strictly less than that of the baseline approach.
Formally, for a grouping plan with $N$ nodes and $k$ groups, each node sends messages to and receives messages from all other $N-1$ nodes, resulting in  transmissions per node:
\begin{equation}
    C_{baseline} = 2 \times (N-1)
    \label{equation: baseline-1}
\end{equation}
In our grouping-based \name{} scheme, each \textbf{Simple Node} communicates only with its aggregator, sending and receiving one message, totaling $2$ transmissions. 
Each \textbf{Aggregation Node} receives and sends messages to group members, and also communicates with the other $k-1$ aggregation nodes. 
In an $N$-node environment, 
the number of group members cannot exceed $ N-k$. Therefore, the total communication rounds is bounded by:
\begin{equation}
    C_{GeoCoCo} \leq 2 \times [(N-k)+(k-1)] = 2 \times (N-1)
    \label{equation: round-guarantee}
\end{equation}
Combining Equation~(\ref{equation: baseline-1}) and Equation~(\ref{equation: round-guarantee}), we obtain $C_{GeoCoCo} \leq C_{baseline}$, which guarantees that even in the worst case, all nodes in \name{} require no more communication rounds than the baseline.




\vspace{0.5ex}
\noindent
\rthree{
\textbf{Aggregator Failover and Fault Tolerance.}
Aggregator failures do not compromise correctness. 
If a relay node fails or becomes unreachable, group members immediately fall back to direct transmission, and the \emph{Grouping Strategy Orchestrator} reconfigures groups in the next round to restore optimized paths.
If a regular node fails, we cannot send or receive data anyway, so we simply skip to the next round and reconfigure the grouping.
Failures are visible to the upper layer, but whether they affect application semantics depends on the application itself; \name{} guarantees that collective transmission among surviving nodes remains correct and consistent. 
CRDT idempotence ensures that retransmissions or duplicates during failover do not cause inconsistency, only potential extra communication or latency. 
These mechanisms are transparent to the database: commit and merge proceed normally, with fewer redundant messages. 
\name{} enforces epoch boundaries to prevent cross-round reordering, keeping each epoch self-contained for validation (e.g., commit or model convergence). Group or aggregator changes across rounds do not affect prior results or future merges.
}

\vspace{0.5ex}
\noindent
\textbf{Reducing Cross-Region Bandwidth Overhead.}
By performing semantic-aware pre-filtering and in-group aggregation, \name{} transmits only essential and non-redundant data across regions. This pre-transmission pruning strategy significantly reduces WAN-level traffic and alleviates communication bottlenecks under limited bandwidth conditions.
Minimizing Latency along Critical Paths
\name{} adopts a hierarchical communication architecture that confines most synchronization operations within low-latency local domains, while limiting inter-region exchange to a small subset of data. It further leverages Triangle Inequality Violation-aware path optimization to circumvent suboptimal links, effectively shortening the end-to-end delay along synchronization-critical paths.
Enhancing global transaction throughput by combining intelligent grouping with redundancy elimination mechanisms, \name{} alleviates the classic “slowest-link bottleneck” problem inherent in fully-connected communication schemes. This design improves the overall transaction processing efficiency in geo-distributed deployments without incurring additional synchronization overhead or compromising consistency guarantees.

\section{IMPLEMENTATION}
We implement the protocol components of \name{} 
with C++ and Python, and integrate them into GeoGauss and CRDB to build our prototype system. This section presents the details.

\vspace{0.2ex}
\noindent
\textbf{Delay Monitoring.}
In \name{}, each replica launches a lightweight background thread during system initialization, which periodically probes link RTTs and reports them to a designated coordinator (the \emph{Monitor}).
The Monitor role can be assigned to an existing worker or to a dedicated node, depending on deployment constraints. 
The probing runs asynchronously and does not block normal transaction execution. 
Since WAN conditions tend to shift in \emph{episodic patterns} rather than continuous fluctuations, \name{} only updates transmission plans when significant and persistent changes are detected, rather than reacting to transient jitter.
This design avoids frequent plan churn while maintaining responsiveness to meaningful network events.
\rfour{
Additionally, to remain scalable when the cluster grows to thousands of nodes—where maintaining a full $N{\times}N$ latency matrix becomes increasingly costly, we employ network coordinate techniques~\cite{ledlie2007network,chatziliadis2024efficient} (e.g., Vivaldi~\cite{dabek2004vivaldi}) to approximate inter-node latencies. 
To ensure estimation accuracy, we integrate a verification mechanism that periodically samples and corrects deviations between predicted and measured RTTs. 
This hybrid design effectively reduces monitoring traffic and computation overhead while maintaining latency fidelity at scale.
}




\vspace{0.2ex}
\noindent
\textbf{LP Solver.}
We utilize Gurobi~\cite{gurobi} to solve our linear programming~(LP) problems.
Gurobi is a commercial optimization solver developed by Gurobi Optimization, LLC, that has been extensively utilized in both academia and industry~\cite{10.14778/3636218.3636227, liu2024rethinking, xu2023teal}. 
It provides state-of-the-art algorithms for solving linear programming, mixed-integer programming (MIP), quadratic programming (QP), and other mathematical optimization problems.
Gurobi employs a \textit{concurrent optimization strategy}, running both \textit{dual simplex} and \textit{barrier methods} in parallel for linear programming.
\rfour{
In theory, for convex LPs solved with barrier methods, the complexity of finding an $\varepsilon$-approximate solution is bounded by $\mathcal{O}(\sqrt{M} \cdot \ln(V/\varepsilon))$ Newton iterations (where $M$ is the condition number, $V$ the initial duality gap, and $\varepsilon$ the accuracy tolerance), where each requires $\mathcal{O}(n^3)$ operations per step ($n$ = problem dimension)~\cite{gurobi_doc, boyd2004convex}.
In our experimental environment (Aliyun g7ne instance), Gurobi can solve a 15-node LP-based grouping problem in less than 10 ms.
This ultra-low computational delay enables us to generate latency-aware transmission topologies in real time, making our dynamic grouping strategy practically deployable.
}

\vspace{0.5ex}
\noindent
\rfour{
\textbf{K-Center–Based Scalable Planner.}
To further improve scalability when the number of nodes grows large, we replace the LP-based grouping in GeoCoCo with a lightweight K-center–based heuristic.
This algorithm aims to minimize the maximum intra-group latency by iteratively selecting the most distant node (in terms of network delay) as a new group center, and assigning all remaining nodes to their nearest center.
By doing so, it approximates the optimal latency-aware grouping with only O($N \times k$) complexity, while guaranteeing that the resulting maximum intra-group delay is within twice of the theoretical optimum.
Compared with the LP solver, the K-center method avoids quadratic complexity and enables near–real-time group planning even when the cluster size scales to hundreds or thousands of nodes.
This heuristic thus preserves GeoCoCo’s latency-awareness and correctness, while substantially reducing the computational cost of planning.
}

\vspace{0.5ex}
\noindent
\rtwo{
\textbf{Overlay-based Implementation.}
\name{} exploits Triangle Inequality Violations without modifying Border Gateway Protocol (BGP) or any low-level routing. 
Instead, it realizes indirect paths purely at the application layer through lightweight user-space relays 
(e.g., TCP port-forwarding or gRPC proxies) deployed on ordinary cloud VMs or existing replicas. 
Relay nodes are selected dynamically based on online RTT measurements and health checks, 
and \name{} automatically falls back to the direct path if a relay fails or does not provide sufficient latency gain. 
This overlay-based deployment requires no privileged access and can be easily implemented in public clouds by simply enabling inter-region ports. All experiments in this paper use this deployment strategy in practice.}

\vspace{0.2ex}
\noindent
\textbf{Transactional Isolation.}
To ensure system correctness in the presence of concurrent updates, we enforce a transactional isolation mechanism when modifying transmission plans. Specifically, at the beginning of each synchronization cycle, the system creates a snapshot of the current transmission plan. All transmission operations within that cycle are then executed strictly based on this immutable snapshot, regardless of whether new plans are generated during execution. This design effectively prevents interference between the planning and execution phases, ensuring that ongoing transmissions are not affected by configuration updates.
As a result, the system maintains consistent behavior and predictable semantics even under dynamic reconfiguration. In other words, even if the system undergoes configuration changes at runtime—such as updating transmission plans, switching aggregation nodes, or adjusting scheduling strategies—the actual execution remains correct, ordered, and consistent with expected semantics.

\vspace{0.2ex}
\noindent
\textbf{Collective Communication.}
\label{sub: co-co}
We integrate \name{} as an intermediate communication layer between the database commit protocol and the transport stack. In GeoGauss, this is achieved by replacing direct point-to-point \texttt{ZeroMQ} calls with \name{}’s collective communication APIs (\texttt{AllToAll}, \texttt{Broadcast}, \texttt{AllReduce}, \texttt{Gather}, \texttt{AllGather}). 
By default, GeoGauss uses low-level \texttt{ZeroMQ} primitives (e.g., \texttt{zmq\_send}, \texttt{zmq\_recv})~\cite{zeromq_github, hintjens2013zeromq, geogauss}, tightly coupling database logic with transport behavior and limiting WAN adaptability. \name{} instead provides a high-level, intent-driven interface: the database specifies the communication pattern, and \name{} automatically selects optimal WAN routing and execution via topology-aware grouping, relay scheduling, and filtering.
This design preserves the architecture of GeoGauss and only replaces the network invocation path. Collective operations (e.g., epoch-based update dissemination) become a single \texttt{geococo.all\_to\_all()} call, enabling automatic scheduling and WAN optimization. 
As a result, \name{} improves WAN performance and can be easily ported to other geo-distributed databases via its general-purpose API.

\vspace{0.3ex}
\noindent 
\textbf{Extensions -- Integration with CockroachDB.}
To demonstrate the generality of \name{}, we discuss its applicability to Raft-based geo-replicated databases such as CockroachDB.
Since \name{} operates at the communication layer, it can be seamlessly integrated with Raft’s transport subsystem without modifying the consensus protocol itself.
Specifically, \name{} can hook into \emph{RaftTransport} to intercept replication messages (e.g., \emph{AppendEntries, Heartbeats}, and \emph{Snapshots}).
The Planner and Communicator modules enable latency-aware grouping and hierarchical forwarding of these messages, while the Filter module can be applied to replication streams to remove redundant updates before WAN transmission. This design reduces cross-region latency and bandwidth consumption while preserving Raft’s quorum semantics.

\section{EVALUATION}

\subsection{Setup}
\label{subsec: setup}

\vspace{0.2ex}
\noindent 
\noindent\textbf{Baselines.}
To contextualize our gains, we evaluate GeoCoCo against both native database systems and representative synchronization optimization techniques. %
Specifically, we reference the following three mainstream directions: %
\begin{itemize}[leftmargin=*]
\item \textbf{GeoGauss}~\cite{geogauss}:
We use the default GeoGauss system as the baseline, which performs full all-to-all synchronization across replicas without grouping or filtering. %
\item \textbf{CockroachDB}~\cite{cockroachdb_latency, taft2020cockroachdb}:
We also compare against unmodified CockroachDB, which relies on its native Raft replication pipeline and direct WAN message delivery. %
\item \textbf{Other Representative Techniques:}
(i) congestion control protocols such as \textbf{BBR}~\cite{cardwell2016bbr}, %
(ii) compression applied to replication streams such as \textbf{zlib}~\cite{zlib,zlib_github}, and %
(iii) scheduling and locality-aware execution strategies exemplified by \textit{Calvin} and \textbf{SLOG}~\cite{ren2019slog}. %
\item \textbf{Grouping Strategies:}
To evaluate our latency-aware grouping mechanism, we compare the cost and per-round performance of our LP-based strategy against several alternatives, including hierarchical agglomerative clustering~\cite{yao2002cougar}, KMeans with 2 and 3 clusters~\cite{lloyd1982least,pedregosa2011scikit}, random grouping, and a no-grouping baseline. %
\end{itemize}

\vspace{0.2ex}
\noindent 
\textbf{Workload.}
We use a widely adopted online transaction processing~(OLTP) benchmark in database systems, Transaction Processing Performance Council benchmark-C~(TPC-C)~\cite{tpcc} and Yahoo! Cloud Serving Benchmark(YCSB)~\cite{cooper2010benchmarking}.

\vspace{0.2ex}
\noindent $\bullet$ \textbf{TPC-C:}
To evaluate the system under varying transactional patterns and communication demands, we design four representative TPCC-style workloads—TPCC-A, TPCC-B, TPCC-C, and TPCC-D—corresponding to write-intensive, read-intensive, balanced, and real-time scenarios, respectively. 
While the official TPC-C benchmark defines a fixed ratio of five transaction types (NewOrder, Payment, OrderStatus, Delivery, and StockLevel), it is a common and accepted practice in prior work (e.g., Calvin~\cite{thomson2012calvin}, GeoGauss~\cite{geogauss}) to customize this distribution to reflect real-world workloads or highlight system behaviors.
(1) TPCC-A (Write-Intensive): Dominated by NewOrder and Payment (>90\%), simulating peak write-heavy scenarios.
(2) TPCC-B (Read-Intensive): Skewed toward OrderStatus and StockLevel, modeling frequent backend queries.
(3) TPCC-C (Balanced): Even distribution of all five transactions, serving as a standard OLTP baseline.
(4) TPCC-D (Real-Time): Emphasizes OrderStatus for responsive user interactions, with moderate write components.
In addition, we follow standard TPC-C performance metrics to measure throughput: 
\textbf{tpmC} (transactions per minute C) represents the number of committed \texttt{NewOrder} transactions per minute, which is the primary official performance indicator of TPC-C and reflects core write throughput. 
\textbf{tpmTotal} represents the total number of transactions of all five types processed per minute, 
providing a broader view of the system’s overall transactional capacity under mixed workloads.

\vspace{0.2ex}
\noindent $\bullet$ \textbf{YCSB:}
YCSB is one of the most popular key-value transaction benchmarks, encompassing insert, delete, and update operations on key-value pairs. 
We synthesized datasets with varying conflict rates using YCSB to validate \name{}'s performance under different contention levels.
We adjust the conflict rate by tuning the $\theta$  parameter in the Zipfian distribution, which controls the key access skew.
Such configurations are widely used in OLTP system evaluations and reflect typical access patterns in geo-distributed deployments.

\vspace{0.8ex}
\noindent 
\textbf{Cluster Setup.} We evaluate \name{} using both emulated and real-world geo-distributed environments.

\vspace{0.4ex}
\noindent$\bullet$ \textbf{Trace-driven Simulation:}
We emulate WAN conditions using local servers and the TC \textit{netem} module~\cite{tc}, guided by real-time link delay traces from Amazon AWS~\cite{aws}. To capture dynamic behaviors such as delay spikes and jitter, we apply Piecewise Cubic Hermite Interpolating Polynomial fitting~\cite{fritsch1980monotone} to AWS latency data, preserving both mean and distributional properties. This fitting process produces over 10,000 synthetic 10$\times$10 delay matrices, which are replayed via TC~\cite{tc} to simulate time-varying inter-node delays. Real-time monitoring ensures emulation accuracy. This setup enables faithful simulation of WAN environments for evaluating the performance and robustness of geo-distributed databases.

\vspace{0.4ex}
\noindent $\bullet$ \textbf{Real-world Testbed Deployment:}
We further deploy a 5-node geo-distributed database across multiple geographic regions to evaluate performance under real-world production conditions. The deployment includes Kalgan, Hohhot, and Hong Kong. Each server is provisioned as an Alibaba Cloud c5 instance, configured with 24 vCPU  48.0 GiB of RAM, and running CentOS 7.6 Server (64-bit)~\cite{alibabacloud2023ecs}.

\vspace{0.4ex}
\noindent $\bullet$ \textbf{Bandwidth Utilization Measurement:}
To estimate the actual WAN bandwidth consumption during system execution, we collect network interface statistics via the Linux /proc~\cite{linuxprocfs} filesystem. This kernel-level mechanism enables direct access to per-interface traffic counters, including outbound data sent through the NIC. All measurements are recorded at the NIC level, ensuring an accurate reflection of the actual WAN egress traffic per node.




\begin{figure}[t!]
    \centering
    \includegraphics[width=0.68\linewidth]{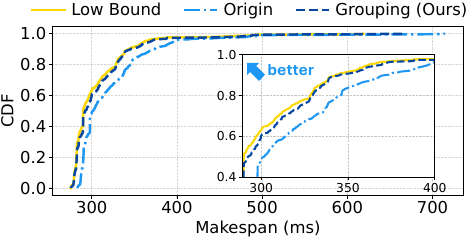}
    \vspace{-0.14cm}
    \caption{CDF Distribution of Single-Round Makespan.}
    \label{fig: cdf}
\vspace{-0.1cm}
\end{figure}

\vspace{0.4ex}
\noindent
\subsection{Micro Benchmark}
\textbf{Theoretical Performance Metric.}
\rone{
We define the completion time of a single-round all-to-all transmission \textit{Makespan} as the system's theoretical performance metric.}
All-to-all is one of the most latency-sensitive and bandwidth-intensive collective communication patterns in geo-distributed environments. Due to network heterogeneity, even if most node pairs complete their transmissions quickly, a few high-latency links can dominate the overall round time, leading to a bottleneck effect. In such systems, the global progress is often gated by the slowest transmission pair, making the Makespan a natural and critical indicator of system efficiency.
Furthermore, in many distributed systems (e.g., databases, machine learning frameworks), communication rounds serve as consistency or synchronization boundaries. Optimizing across rounds risks breaking correctness guarantees. Therefore, we focus on \textit{per-round performance} and avoid cross-round reordering or pipelining.
Formally, we aim to minimize the worst-case transmission latency in a round:
\begin{equation}
\min T_{\text{Makespan}} = \min\left( \max_{i,j \in \{1,2,\dots,N\}} T_{i,j} \right)
\label{equation-min}
\end{equation}

\noindent This objective reflects our design principle: achieving global efficiency by eliminating the longest delays that constrain system progress.

\begin{figure}
\vspace{-0.2cm}
    \centering
    \includegraphics[width=0.65\linewidth]{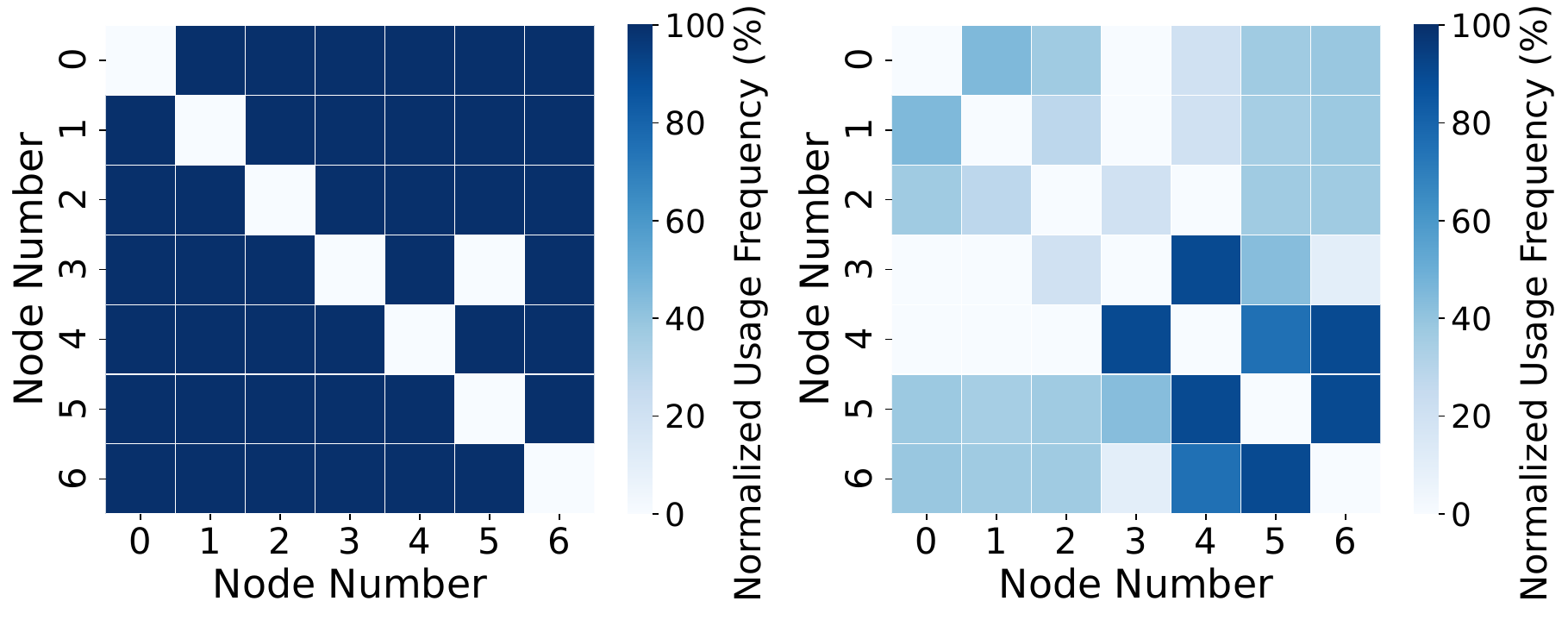}
    \caption{(a) Heatmap of Communication Frequency Without Grouping (Baseline). (b) Heatmap of Communication Frequency with Transmission Grouping Using \name.}
    \label{fig: heatmaps}
\vspace{-0.4cm}
\end{figure}


\vspace{0.4ex}
\noindent 
\textbf{Single-Round Makespan Reduction.}
Figure~\ref{fig: cdf} shows the CDF of single-round all-to-all makespan under three schemes: \textit{Origin} (default), \textit{Grouping (Ours)}, and \textit{Low Bound} (theoretical optimum). The x-axis denotes makespan (ms), and the y-axis shows the CDF across runs.
Our grouping consistently outperforms the baseline: the CDF curve shifts left, indicating more rounds finishing at lower latency. At the 90th percentile, our approach reduces makespan by over 100\,ms. The inset (300–400\,ms range) shows a tighter spread and lower tail latency, moving closer to the theoretical bound.
These gains stem from latency-aware LP grouping, which forms compact groups to bypass slow links and reduce long-path dependencies. While \textit{Origin} suffers from scattered worst-case latency due to flat all-to-all communication, \name{} mitigates bottlenecks from outlier nodes, reducing variance and approaching the optimal frontier.
Overall, \name{} significantly accelerates single-round synchronization, reduces variance, and approaches the theoretical minimum, validating our objective in Equation~(\ref{equation-min}).

\vspace{0.4ex}
\noindent \textbf{Cutting Communication Rounds via Hierarchical Design.} We conducted a communication frequency profiling experiment across 400 rounds of all-to-all transmission over 7 nodes, under a trace-driven simulated WAN latency environment. Figure~\ref{fig: heatmaps} shows the normalized heatmap of point-to-point communication frequency. Compared to the baseline scheme (left, (a)), \name{} (right, (b)) significantly reduces the number of communication rounds for most nodes. Post-hoc inspection reveals that communication is concentrated on a few aggregation nodes (e.g., nodes 3, 4, 5, and 6). Although these aggregation nodes handle more relay traffic, their total communication count remains below that of any single node in the baseline. 

\subsection{Macro Benchmark}

\begin{figure}
\vspace{-0.1cm}
    \centering
    \includegraphics[width=0.7\linewidth]{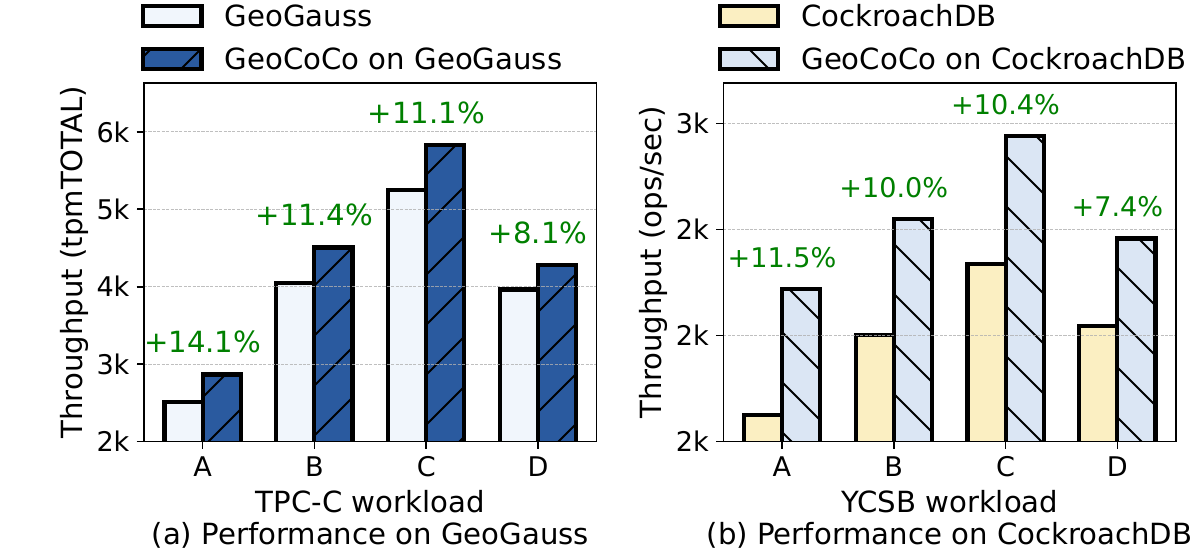}
    \vspace{-0.1cm}
    \caption{Throughput improvement of \name{} (a) over GeoGauss under TPC-C workload A-D and (b) over CockroachDB under YCSB workloads A–D.}
    \label{fig: end-to-end}
\vspace{-0.2cm}
\end{figure}

\textbf{End-to-end Benefits on GeoGauss}.
To evaluate the end-to-end benefits of \name{} in a real database system, we integrate it into GeoGauss, a full-replica multi-master DBMS, and run TPC-C workloads~\cite{tpcc}.
Following the deployment in \S\ref{subsec: setup}, we deploy GeoGauss across 5 geo-distributed data centers 
(two in Kalgan, two in Hohhot, and one in Hong Kong). 
We compare (1) GeoGauss with \name{} optimizations and (2) the default system with native messaging.
To stress inter-node coordination and highlight communication effects, we run 100 TPC-C warehouses, ensuring frequent cross-site updates. Both systems run to a steady state, after which we collect metrics over a sustained sampling window for statistical reliability.
Figure~\ref{fig: end-to-end}(a) presents the average transaction throughput (tpmTOTAL) under four representative TPC-C workloads. We make the following observations:
(1) Consistent Performance Gains: Across all four workloads—TPCC-A,B,C and D)—the system integrated with \name{} consistently outperforms the baseline GeoGauss.
(2) Significant Improvement in Write-Intensive Scenarios: In the TPCC-A workload, which involves frequent cross-node write operations, \name{} achieves the highest improvement of 14.1\%, indicating its ability to mitigate communication bottlenecks in high-contention settings.
(3) Broad Applicability Across Workload Types: Even in read-intensive (TPCC-B), balanced (TPCC-C), and real-time update (TPCC-D) workloads, \name{} delivers non-trivial throughput improvements ranging from 8.1\% to 11.4\%, highlighting its general effectiveness across diverse OLTP scenarios.


\vspace{0.4ex}
\noindent
\textbf{End-to-end Benefit on CockroachDB(CRDB).}
\name{} is integrated into CockroachDB’s Raft-based replication stack by hooking into the RaftTransport layer to intercept replication messages, without modifying the Raft consensus protocol; this non-intrusive design preserves standard quorum semantics. We evaluated the integrated system under YCSB workloads A–D in a real-world wide-area testbed arranged in a full-mesh topology. 
Figure~\ref{fig: end-to-end}(b) shows the throughput for these workloads, comparing baseline CockroachDB with CockroachDB+\name{}. \name{} improves throughput by up to 11.5\% to the baseline (reduces p99 latency by 14.49\% at the same time).
These results demonstrate that \name{}’s latency-aware coordination remains effective in practical Raft-based systems such as CockroachDB.

\subsection{\name{} Deep Dive}
\label{sub: deep-dive}

\begin{figure}
    \centering
    \vspace{-0.2cm}
    \includegraphics[width=0.7\linewidth]{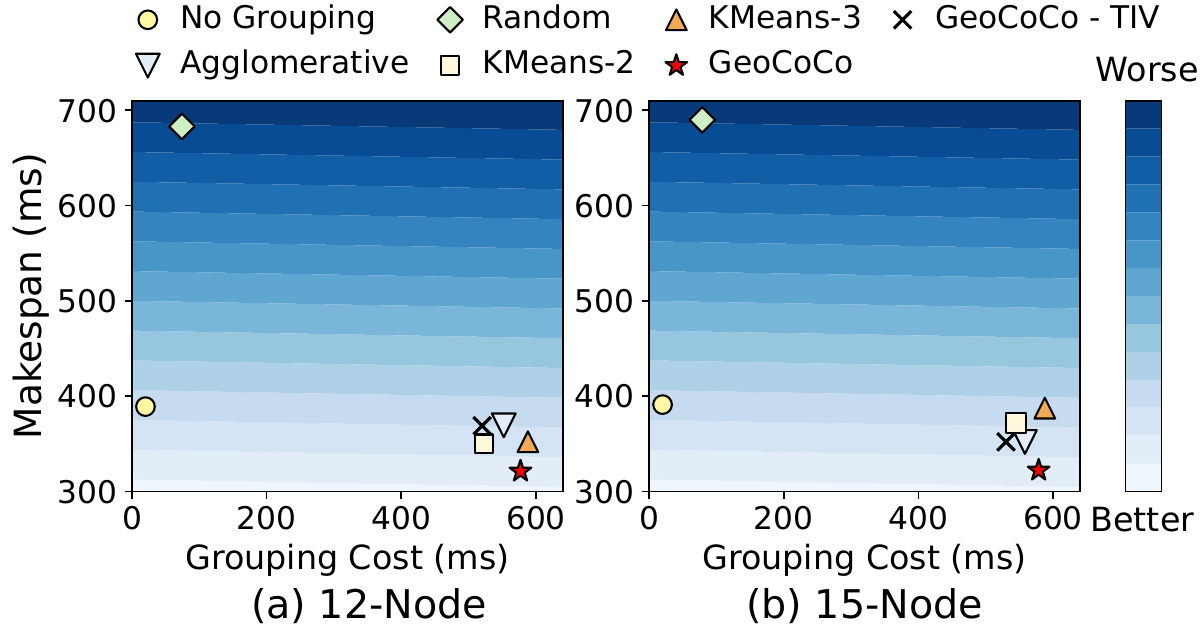}
    \vspace{-0.1cm}
    \caption{Grouping cost vs. communication efficiency under 12-node (a) and 15-node (b) settings. Lighter areas with an arrow indicate better trade-offs.}
    \label{fig: cost}
\vspace{-0.1cm}
\end{figure}

\vspace{0.5ex}
\noindent
\textbf{Comparison of Different Grouping Strategies.}
We evaluate the cost and per-round performance of our Linear Programming (LP)-based grouping scheme against several alternatives, including hierarchical agglomerative clustering~\cite{yao2002cougar}, KMeans with 2 and 3 clusters~\cite{lloyd1982least, pedregosa2011scikit}, random grouping, and a no-grouping baseline. 
\rthree{
To assess the contribution of Triangle Inequality Violations, we additionally evaluate a grouping variant with TIV support disabled, 
enabling an isolated measurement of the benefit brought by TIV-aware grouping(in Figure~\ref{fig: cost}'s GeoCoCo-TIV).}
The x-axis represents the average computation time per grouping, while the y-axis shows the makespan of a single transmission round as shown in Figure~\ref{fig: cost}. The contour lines indicate the trade-off boundary where the grouping plan is updated every 10 rounds.
We highlight 4 key observations:
\textbf{First}, \name{}'s LP-based method consistently outperforms all baselines under both 12-node and 15-node settings, demonstrating strong generality and stability across scales.
\rfour{
\textbf{Second}, although \name{} incurs higher computational overhead as the cluster size grows, its performance gains scale accordingly.
For example, with 12 nodes, \name{} achieves a 16.46\% improvement in makespan, which is sufficient to offset the cost within a single synchronization round.
When the number of nodes increases to 15, the solving time rises (comparable to KMeans and hierarchical agglomerative clustering), but \name{} yields a 17.63\% improvement—significantly higher than the best baseline (11.0\% for hierarchical clustering). This makes the additional computation cost easily amortized within just a few rounds, and because grouping runs asynchronously, the overhead does not block ongoing execution and is fully offset after only a small number of synchronization rounds.
}
\rthree{
\textbf{Third}, Enabling TIV grouping (Full \name{}) reduces makespan by 7.62\%–12.44\% beyond grouping alone, demonstrating that TIV exploitation offers an independent and additive benefit.
}
\textbf{Last}, \name{} consistently provides performance benefits. This is because it avoids frequent regrouping and only updates plans when stable, ensuring sustained efficiency over time.



\begin{figure}[t]
    \centering
    \begin{minipage}{0.36\linewidth}
        \includegraphics[width=\textwidth]{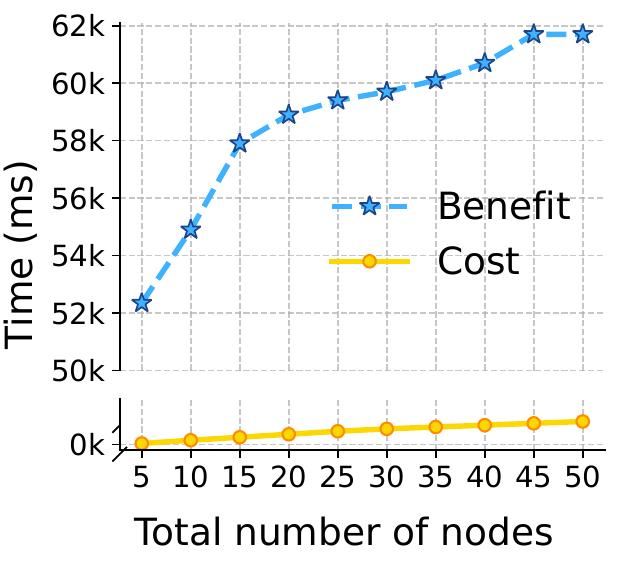}
        \vspace{-0.42cm}
        \caption{Comparison of grouping cost and cumulative benefit of \name{} under increasing cluster sizes.}
        \label{fig: big-cluster}
    \end{minipage}
    \hspace{0.04\linewidth}
    \begin{minipage}{0.36\linewidth}
        \vspace{-0.26cm}
        \includegraphics[width=\textwidth]{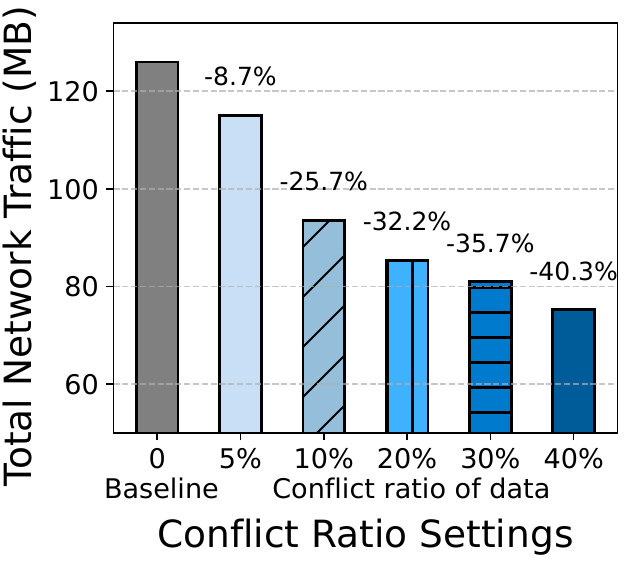}
        \vspace{-0.42cm}
        \caption{Bandwidth Reduction under Conflict-Aware Aggregation.}
        \label{fig: aggregation}
    \vspace{0.35cm}
    \end{minipage}
\vspace{-0.4cm}
\end{figure}

\vspace{0.8ex}
\noindent
\textbf{Cost and Benefit in Larger Cluster.}
We systematically evaluate the scalability of \name{} through trace-driven simulation experiments with node counts ranging from 5 to 50. 
The 50-node deployment represents a geo-distributed setup spanning multiple continents and regions, exceeding the largest data-center-scale configurations observed in existing geo-distributed systems—25 nodes in PigPaxos~\cite{charapko2021pigpaxos} and 48 nodes across 3 AZs in CRDB~\cite{taft2020cockroachdb}.
Figure~\ref{fig: big-cluster} compares \name{} with the original non-hierarchical transmission scheme. The Cost curve (yellow circles) shows the one-time overhead of computing the hierarchical plan, while the Benefit curve (blue stars) shows the cumulative reduction in synchronization time over 1000 rounds. Each round is triggered every 10,ms, consistent with the default GeoGauss setting.
Although planning overhead increases with cluster size, the cumulative benefit rapidly exceeds the cost. 
From 5 to 50 nodes, the planner’s cost accounts for only 6.65\% to 7.07\%.
This is enabled by our guided group search strategy, which reduces planning complexity by about an order of magnitude relative to exhaustive LP-based search across all group counts in $[2, N{-}1]$. 
In addition, planning occurs before data transmission and can be offloaded to an idle node, avoiding interference with runtime execution. 
Overall, \name{} scales efficiently to large clusters, providing substantial synchronization gains at minimal amortized planning cost.

\vspace{0.3ex}
\noindent
\textbf{Bandwidth Efficiency and Filtering Overhead.}
\rone{
To assess the impact of aggregation on bandwidth utilization, we conduct controlled experiments using the YCSB benchmark with a total of 1,000,000 operations.
}
\rtwo{
As shown in Figure~\ref{fig: aggregation}, the baseline (gray bar) represents no filtering, while the remaining bars correspond to conflict ratios of 5\%–40\%.
The y-axis measures the total network traffic across all nodes in the system (in MB), which reflects the overall bandwidth consumption. 
The total network traffic across all nodes decreases consistently as the conflict ratio increases—by 8.7\%, 27.2\%, 32.2\%, 35.7\%, and 40.3\%, respectively.
}
This monotonic reduction confirms that higher contention introduces more 'white' data (this trend aligns with Observation~\#2), which filtering effectively eliminates to reduce WAN transmission costs.
\dl{
We further tested an extreme conflict-free case (randomized primary key reads). In this setting, the filtering phase is effectively bypassed, and the system’s performance remains within 1\% of the grouping-only baseline. This confirms that the filtering mechanism is lightweight and imposes negligible cost when no redundant writes exist.
}
\dl{
To quantify the runtime cost of white-data identification under normal workloads, we profile CPU and tail latency with filtering enabled and disabled. The detection path performs O(1) in-memory version checks and conflict hints with no additional coordination. 
Table~\ref{tab:filter-overhead} illustrates that across YCSB workloads, filtering incurs <2.76\% CPU and < 9.22 ms change in p99 latency (excluding the benefits of grouping), while reducing cross-region traffic by up to 40.3\%.
These results confirm that filtering is lightweight and non-intrusive, and does not form a bottleneck even under high load.
}

\begin{table}
\centering
\small
\caption{Filtering overhead and WAN savings across conflict ratios (YCSB).}
\label{tab:filter-overhead}
\vspace{0.3cm}
\begin{tabular}{@{}lcccc@{}}
\toprule
\textbf{Conflict Ratio} & \textbf{Config} & \textbf{CPU↑} & \textbf{p99 (ms)} & \textbf{WAN↓} \\
\midrule
\multirow{2}{*}{0\%}
  & Without Filtering & --      & --     & --   \\
  & With Filtering    & +0.97\% & +0.8   & $\sim$0\% \\
\midrule
\multirow{2}{*}{30\%}
  & Without Filtering & --      & --     & --   \\
  & With Filtering    & +2.41\% & +8.4   & 35.7\% \\
\midrule
\multirow{2}{*}{40\%}
  & Without Filtering & --      & --        & --   \\
  & With Filtering    & +2.76\% & +13.22    & 40.3\% \\
\bottomrule
\end{tabular}
\end{table}

\vspace{0.4ex}
\noindent
\rfour{
\textbf{Cost of Delay Monitoring.}
We measure the runtime overhead of delay monitoring under different cluster scales.
In our standard configuration, even with per-epoch high-frequency probing on a 50-node full mesh, the total monitoring traffic is only about 5.2 MB/s (0.1 MB/s per node) with 0.7–1.1\% CPU utilization, indicating negligible runtime impact at moderate scales.
\rone{
To further ensure scalability at very large scales, we employ a \emph{Vivaldi-based network coordinate system} (NCS) to estimate inter-node latencies instead of performing full pairwise measurements.}
Under a 1,024-node setting using real-world latency traces, the NCS-based approach reduces detection overhead by 96.4\% while maintaining estimation accuracy within 18\% of direct probing, demonstrating its practicality for large-scale deployments.
}

\vspace{-.1in}
\dl{
\subsection{Combining with other work}
Many studies~\cite{cardwell2016bbr, zlib, zlib_github, ren2019slog, thomson2012calvin} 
accelerate geo-distributed databases by optimizing transaction scheduling, improving network efficiency, or reducing data volume via compression.
\name{} targets synchronization efficiency, orthogonal to these directions.
We further evaluate its performance when combined with such optimizations.}

\begin{figure}[t]
\vspace{0.1cm}
    \centering
    \begin{minipage}{0.32\linewidth}
    \vspace{-0.06cm}
        \includegraphics[width=\textwidth]{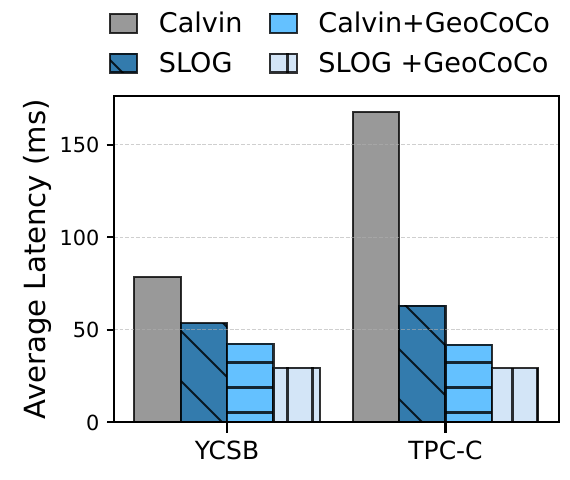}
        \vspace{-0.62cm}
        \caption{Combining GeoCoCo with Calvin and SLOG under YCSB and TPC-C.}
        \label{fig: schedule}
    \end{minipage}
    \hspace{0.08\linewidth}
    \begin{minipage}{0.32\linewidth}
        \includegraphics[width=\textwidth]{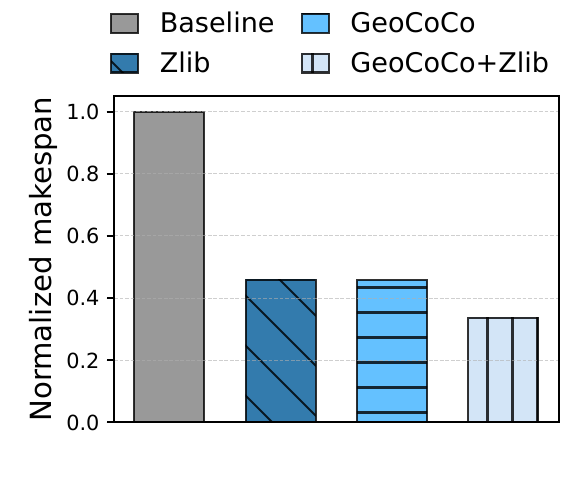}
        \vspace{-0.6cm}
        \caption{Makespan under compression baseline (Zlib) and GeoCoCo (normalized).}
        \label{fig: compression}
    \end{minipage}
\end{figure}

\vspace{0.3ex}
\noindent
\dl{
\textbf{Combining with Other Transaction and Synchronization Scheduling Protocols.}
We evaluate \name{} together with two representative deterministic and locality-aware \textbf{execution protocols}, Calvin~\cite{thomson2012calvin} and SLOG~\cite{ren2019slog}.
Both systems deterministically coordinate cross-partition transactions and employ 
execution scheduling to mitigate wide-area delays.
We integrate \name{} into their communication layer without modifying their transaction ordering logic or guarantees. 
As shown in Figure~\ref{fig: schedule}, \name{} consistently lowers end-to-end latency on both YCSB and TPC-C workloads.
When integrated with Calvin, GeoCoCo reduces average latency by 45.95\%, and by 53.53\% when integrated with SLOG.
This result indicates that GeoCoCo can achieve performance benefits under different execution protocols.
}

\vspace{0.2ex}
\noindent
\rtwo{
\textbf{Combining with Compression Techniques.}
Compression is a widely used technique for reducing data volume.
We select zlib~\cite{zlib, zlib_github}, one of the most popular compression libraries, to evaluate the performance of integrating compression with \name{}.
We conducted experiments on the YCSB workload, where the data transmission block size was set to 4 MB.
Figure~\ref{fig: compression} reports the normalized makespan of one synchronization round across four configurations: Baseline, zlib, \name{}, and \name{}+zlib. 
Compression alone reduces the makespan by 54.13\% compared to the baseline by shrinking data volume, while \name{} achieves a larger reduction through WAN-aware path planning and semantic filtering. 
Combining zlib with \name{} yields the lowest makespan, which is 33.62\% of the baseline.
The results show that byte-level compression and \name{} target complementary dimensions and can be effectively stacked.
}

\begin{figure}
\vspace{-0.2cm}
    \centering
    \includegraphics[width=0.72\linewidth]{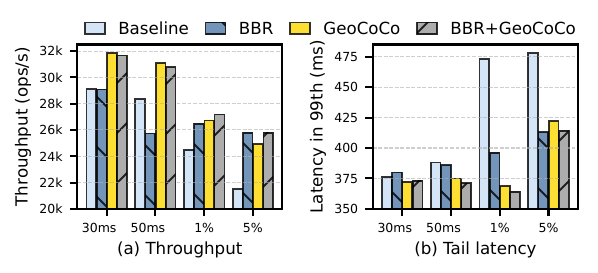}
    \vspace{-0.6cm}
    \caption{\textbf{WAN loss and jitter robustness with BBR comparison.}
    (a) shows throughput under packet loss (1\%, 5\%) and RTT jitter
    (+30\,ms, +50\,ms) across 4 configurations. (b) shows the corresponding p99 latency.}
    \label{fig: trans}
\vspace{-0.2cm}
\end{figure}

\vspace{0.2ex}
\noindent
\dl{
\textbf{Combining with Advanced Network Protocols.}
A variety of network transmission protocols have been developed to address the challenges of network loss and jitter in wide-area settings.
BBR~\cite{cardwell2016bbr} is a representative work, which estimates the bottleneck bandwidth and round-trip propagation delay to pace traffic at an optimal rate, achieving high throughput while maintaining low latency.
We evaluate the performance of combining \name{} with BBR under different network conditions.
In practical, we extend the trace-driven setup
from \S \ref{subsec: setup} with controlled network impairments using Linux \texttt{tc netem}.
We inject (i) packet loss at 1\% and 5\%, and (ii) RTT inflation of +30\,ms and +50\,ms to emulate WAN jitter while preserving the original latency distribution.
We measure throughput, p99 latency, and single-round makespan.
As shown in Figure~\ref{fig: trans}, we summarize the results as follows.
(1) Under 1\% and 5\% packet loss, \name{} sustains higher throughput (by 9.3\%–15.8\%) and reduces p99 latency by 56–104,ms (up to 22\%) relative to the Baseline, demonstrating effective handling of loss-induced retransmissions and WAN uncertainties.
(2) With +30,ms and +50,ms RTT inflation, \name{} consistently lowers tail latency (by 4–13,ms) and improves throughput by 9.3\%–9.6\%, indicating strong robustness to time-varying WAN delays.
}
\rtwo{
This indicates that \name{} provides robustness complementary to transport-layer congestion control, continuing to deliver performance benefits even in lossy and jitter-heavy WAN environments.
}

\subsection{Discussion}

\vspace{0.3ex}
\noindent

\begin{figure}
\vspace{0.06cm}
    \centering
    \includegraphics[width=0.7\linewidth]{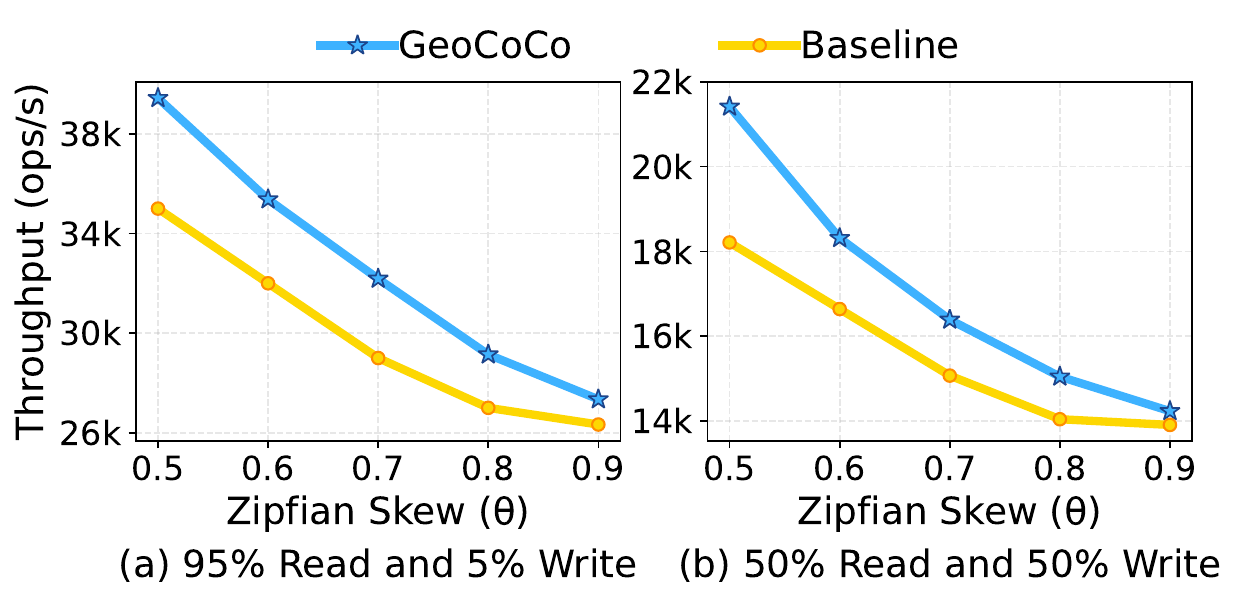}
    \vspace{-0.28cm}
    \caption{Impact of Access Skew on Throughput Under Different Read/Write Ratios.}
    \label{fig: zipf}
\vspace{-0.3cm}
\end{figure}

\vspace{0.3ex}
\noindent
\rfour{
\textbf{Impact of Access Skew (Zipfian).}
\label{sub:zipf}
We vary the Zipfian skew parameter $\theta$ from $0.5$ to $0.9$ to emulate increasingly hotspot-dominated workloads. Figure~\ref{fig: zipf} reports results under both read-dominant (95\% read) and balanced (50\% read) settings. Under the 95/5 workload, \name{} improves throughput over the baseline by 12.7\%, 10.5\%, 10.9\%, and 7.9\% at $\theta{=}0.5$, $0.6$, $0.7$, and $0.8$, respectively, and maintains similar performance at extreme skew ($\theta{=}0.9$; 27.3k vs. 26.3k ops/s).
Under the 50/50 workload, the benefit persists: GeoCoCo delivers 17.6\%, 10.0\%, 8.7\%, and 7.2\% higher throughput at $\theta{=}0.5$–$0.8$, and remains slightly ahead even at $\theta{=}0.9$ (14.2k vs. 13.9k ops/s).
These results confirm that GeoCoCo sustains clear gains across skew levels and read/write ratios, particularly in realistic moderate-skew regimes ($\theta{=}0.5$–$0.8$), where WAN communication remains the dominant bottleneck.
}

\begin{figure}
\vspace{-0.1cm}
    \centering
    \includegraphics[width=0.65\linewidth]{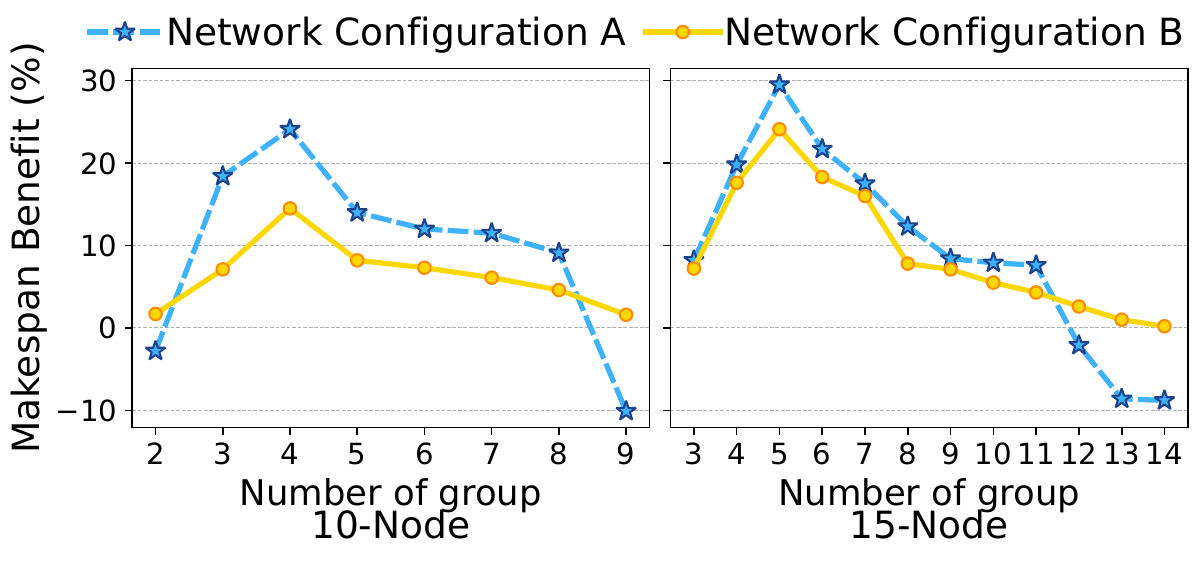}
    \vspace{-0.45cm}
    \caption{Optimal Group Number Analysis for Efficient Geo-Distributed Communication.}
    \label{fig: number-of-group}
\vspace{-0.25cm}
\end{figure}

\vspace{0.3ex}
\noindent
\textbf{The Setting of Group Number.}
\label{sub: group-number}
The number of groups is a key hyperparameter in \name{}, influencing both planning cost and communication efficiency. Figure~\ref{fig: number-of-group} evaluates different group sizes under 10- and 15-node clusters across two WAN settings. The x-axis shows the group count, and the y-axis shows the makespan reduction over no grouping. The theoretical optimum $k^*$ (\S\ref{subsec: grouping}) is derived from expected communication frequency, and the empirical optima (4 and 5 groups) align with our analysis. This illustrates the core trade-off: more groups reduce inter-group cost but increase intra-group coordination. Overall, the results confirm that our cost model effectively guides group selection in practice.

\vspace{0.3ex}
\noindent
\rthree{
\textbf{Inter-group conflicts.}
Our filtering mechanism primarily focuses on intra-group redundancy, as redundant updates (e.g., repeated writes to the same key) are typically concentrated within latency-proximate groups that align with geographic or workload locality.
Although inter-group redundancy (e.g., hot keys or shared counters) can occur, supporting cross-group filtering would require additional global coordination and digest exchanges across aggregators, which could introduce nontrivial overhead and undermine the lightweight nature of \name{}.
Given that intra-group redundancy dominates in realistic geo-distributed workloads, this design strikes a practical balance between effectiveness and complexity, while leaving room for future exploration of global coordination when justified by workload characteristics.
}

\section{RELATED WORK}


Numerous research efforts have been dedicated to enhancing transmission performance in distributed database systems~\cite{huang2025chimera, zhuang2025geotp, li2022alnico, choi2023hydra, jepsen2021network, li2022switchtx, schuh2021xenic, sun2023neobft, song2023network, jin2018netchain, grpc_github, indrasiri2020grpc, rfc9000, rfc9001, Li2025WADNsurvey, zhang2023quic_fastnet, gunther20170, li2020tack, arun2018copa, cardwell2016bbr, xu2015reducing, xu2017online, shilane2012wan, shapiro2011crdt, zhang2021efficient, zhang2023efficient, shenmako, huang2025chimera, zlib, zlib_github}, primarily focusing on 1) cross transaction-transmission scheduling, 2) transmission optimization, 3) data transmission reduction.

\vspace{0.5ex}
\noindent
\textbf{Cross Transaction-Transmission Scheduling.}
The goal of data transmission in a distributed database is to ensure the correctness of transactions.
Considering the hardware feature of network devices, many researchers have dedicated for optimizing the scheduler between transaction and transmission~\cite{li2022alnico, choi2023hydra, jepsen2021network, li2022switchtx, schuh2021xenic, sun2023neobft, xu2025geolm, song2023network, jin2018netchain, zhang2023efficient, shenmako, huang2025chimera, zhuang2025geotp}.
For example, 
AlNiCo~\cite{li2022alnico} leverages FPGA-based smart NICs to schedule transaction requests, thereby reducing contention among multiple CPU cores.
Hydra~\cite{choi2023hydra} uses programmable switches (P4) to perform transaction ordering at the network layer, offloading the ordering logic from servers to reduce their processing burden.
SwitchTx~\cite{li2022switchtx} offloads distributed transaction coordination to a tree of programmable switches. It reorders in-flight transaction messages based on semantic priorities and tightly couples admission control with congestion control to optimize traffic.
Xenic~\cite{schuh2021xenic} serves as a smart-NIC-optimized transaction engine. It adopts an asynchronous aggregation execution model with custom NIC logic to maximize both network and core efficiency, thereby reducing commit latency.
Compared to approaches that integrate transaction processing with network-level scheduling using custom hardware such as programmable switches or smart NICs, \name{} does not rely on any dedicated network devices or modify the underlying transaction scheduling mechanism. This design allows \name{} to be seamlessly integrated with existing concurrency control schemes at minimal integration cost, providing complementary benefits.

\vspace{0.5ex}
\noindent
\textbf{Cross-region transmission optimization.}
Cross-region transmission optimization improves communication efficiency across geographically distributed data centers. Compared to intra-datacenter networks, WANs exhibit higher latency, fluctuating bandwidth, and dynamic link conditions, posing unique challenges for efficient data transfer.
To address these issues, modern systems adopt optimizations across multiple layers~\cite{grpc_github, indrasiri2020grpc, rfc9000, rfc9001, zhang2023quic_fastnet, gunther20170, li2020tack, arun2018copa, cardwell2016bbr, Li2025WADNsurvey}. 
At the application layer, RPC frameworks like gRPC~\cite{grpc_github, indrasiri2020grpc} leverage HTTP/2 to enable long-lived, pipelined streaming. 
At the transport layer, QUIC—built on UDP—integrates TLS~\cite{rfc9000, rfc9001, zhang2023quic_fastnet}, 0-RTT handshake~\cite{gunther20170}, multiplexing, and connection migration, making it well-suited for WAN conditions. 
At the \textbf{network level}, protocols like TACK~\cite{li2020tack} reduce ACK overhead on high-latency paths, while Copa~\cite{arun2018copa} balances latency and fairness in congestion control.
\name{} differs from these designs in two key ways: it integrates real-time latency into dynamic planning and optimizes transmission at a topology level, rather than treating flows in isolation. This enables \name{} to adapt to live network conditions and optimize WAN-wide performance.

\vspace{0.5ex}
\noindent
\rone{
\textbf{Data transmission reduction.}
Reducing network traffic to improve transmission efficiency constitutes a classic research direction~\cite{xu2015reducing, xu2017online, shilane2012wan, shapiro2011crdt, zhang2021efficient, yao2002cougar, madden2005tinydb, shen2001sensor,yuan2020efficient,yao2023ragraph, chatziliadis2024efficient, bartolomeo2023oakestra, zlib, zlib_github, zhang2021tadoc, chen2023compressgraph, chen2025compressgnn, 11152825, xu2024improving}. 
To minimize data transfer, geo-replicated systems typically leverage a combination of compression~\cite{zhang2021efficient}, deduplication~\cite{xu2015reducing,xu2017online}, delta encoding~\cite{shilane2012wan}, and aggregation~\cite{yao2002cougar,madden2005tinydb,shen2001sensor,yuan2020efficient,yao2023ragraph,chatziliadis2024efficient,bartolomeo2023oakestra, 10949833}.
} 
In particular, delta encoding transmits only the differences or compressed change logs rather than full records.
For instance, 
dbDedup~\cite{xu2017online} is an online deduplication scheme for databases that leverages similarity-aware, record-level delta encoding to reduce both storage usage and replication traffic.
PHLIP~\cite{shilane2012wan}, etc. introduces a wide-area data replication architecture that integrates stream-informed delta compression into existing deduplication systems, effectively addressing bandwidth limitations in remote backup scenarios.
Cougar~\cite{yao2002cougar}, etc. pushes operators to intermediate nodes for in-network aggregation, avoiding the transmission of raw readings from all sources and significantly reducing network traffic.
\name{} represents the first work applying these data reduction principles at the network layer of distributed databases. 
Its novel approach significantly reduces network load while strictly maintaining transaction correctness, improving transmission performance by 40.3\% by reducing bandwidth usage.
\vspace{-0.15cm}

\section{CONCLUSION}
This paper presents \name{}, a latency-aware group-based synchronization framework designed to address the inefficiencies of geo-distributed database coordination. By systematically identifying structural opportunities—such as regional aggregation hubs, speculative low-latency triangular paths, and data update redundancy—\name{} rethinks the way synchronization is performed across WANs. 
Our proposed hierarchical group-based transmission design, combined with redundancy-aware filtering and aggregation mechanisms, jointly improves system throughput, reduces overall makespan and transmission cost, while preserving the transactional correctness of data delivery.
Experiments on GeoGauss and CockroachDB, as well as large-scale trace-driven evaluations, demonstrate up to 40.3\% reduction in synchronization cost and 14.1\% throughput improvement, with scalable planning overhead amortized in only a few synchronization rounds. 
These results highlight the importance of topology-aware design in scaling distributed databases across geographically dispersed infrastructures.

\section{ACKNOWLEDGMENTS.}

This work was partly inspired by early-stage ideas that arose from discussions 
with Huawei experts Yang Ren and Jiaqi Liang, and was incubated with support from 
our joint collaboration project with Huawei. We also thank Yunpeng Chai, Zili Meng and Feng Zhang for their valuable guidance.

This work is supported by the National Natural Science Foundation of China (NSFC) 
under Projects No.62202473, No.62572473, 
No.62272466, No.62441230, No.62322213, No.62441230, No.62461146205, No. 62461146205 and No.62461146205, Shuimu Scholar Fellowship of Tsinghua University, as well as by the Postdoctoral Fellowship Program of CPSF under Grant No.GZC20251044.

%

\bibliographystyle{plain}
\bibliography{sample-base}

\end{document}